\documentclass[aps,prd,nofootinbib,showpacs,superscriptaddress, preprint]{revtex4-1}
\pdfoutput=1
\usepackage{amsmath,amsfonts,amssymb,graphicx,natbib,bm}
\usepackage{color}
\usepackage{slashed}    
\usepackage{color}
\usepackage{url}
\usepackage{cancel}
\usepackage{slashed}
\usepackage{multirow}
\usepackage{rotating}
\usepackage{braket}
\usepackage{float}
\usepackage{subfigure}
\usepackage[normalem]{ulem}
\usepackage[dvipsnames]{xcolor}

\usepackage{hyperref}
\hypersetup{
	colorlinks=true,
	linkcolor=red,
	filecolor=magenta,
	urlcolor=magenta,
	citecolor=RoyalBlue}

\newcommand{\be}{\begin{eqnarray*}}
\newcommand{\ee}{\end{eqnarray*}}

\newcommand{\eee}{\end{eqnarray}}
\newcommand{\beeq}{\begin{equation}}
\newcommand{\eeq}{\end{equation}}

\usepackage{xspace}

\newcommand{\ba}{\begin{array}}
\newcommand{\ea}{\end{array}}
\newcommand{\bd}{\begin{displaymath}}
\newcommand{\ed}{\end{displaymath}}
\newcommand{\besub}{\begin{subequations}}
\newcommand{\eesub}{\end{subequations}}
\newcommand{\bea}{\begin{eqnarray}}
\newcommand{\eea}{\end{eqnarray}}

\def\l{\lambda}

\def\D{\Delta}

\def\q2 {q^2}

\def\bt{\begin{table}}
\def\et{\end{table}}

\begin{document}

\title{Scalar Triplet Flavor Leptogenesis with Dark Matter}

\author{Arghyajit Datta}
\email{datta176121017@iitg.ac.in}
\affiliation{Department of Physics, Indian Institute of Technology Guwahati, Assam 781039, India}

\author{Rishav Roshan}
\email{rishav@prl.res.in}
\affiliation{Physical Research Laboratory, Ahmedabad - 380009, Gujarat, India}

\author{Arunansu Sil}
\email{asil@iitg.ac.in}
\affiliation{Department of Physics, Indian Institute of Technology Guwahati, Assam 781039, India}

\vspace{1.5 cm}
\begin{abstract}  
We investigate a simple variant of type-II seesaw, responsible for neutrino mass generation, where the particle 
spectrum is extended with one singlet right-handed neutrino and an inert Higgs doublet, both odd under an additional 
$Z_2$ symmetry. While the role of the dark matter is played by the lightest neutral component of the inert Higgs doublet (IHD), 
its interaction with the Standard Model lepton doublets and the right-handed neutrino turns out to be crucial in generating the correct baryon abundance of the Universe through flavored leptogenesis from the decay of the $SU(2)_L$ scalar 
triplet, involved in type-II framework. 
We observe a correlation between the smallness of the mass splitting, among the dark matter and the CP-odd neutral scalar 
from the IHD, and the largeness of the mass of the triplet followed from the dominance of the type-II mechanism over the radiative contribution to neutrino mass.
\end{abstract}

\maketitle
\section{Introduction}

Existence of tiny but non-zero neutrino mass \cite{Fukuda:1998mi,Ahmad:2002jz,Eguchi:2002dm,Ahn:2002up} along with the observed excess of matter over antimatter in the Universe \cite{Riotto:1999yt,Dine:2003ax}
are undoubtedly two of the most challenging  problems in the present day particle physics and cosmology which signal 
for physics beyond the Standard Model (SM). Among many promising scenarios came up as a resolution to these issues, the seesaw mechanism provides an elegant framework to deal with. As is well known, in case of 
type-I seesaw \cite{Minkowski:1977sc,GellMann:1980vs,Mohapatra:1979ia,Schechter:1980gr}, presence of SM singlet right handed neutrinos (RHN) not only helps in generating tiny neutrino mass but they can also be responsible for explaining the observed matter-antimatter asymmetry via leptogenesis~\cite{Fukugita:1986hr,Buchmuller:2004nz,Anisimov:2007mw,Davidson:2008bu,Buchmuller:2005eh,Davoudiasl:2015jja,Narendra:2017uxl,Dolan:2018qpy,Kashiwase:2012xd,Konar:2020vuu,Barman:2021tgt,Bhattacharya:2021jli}. A variant 
of it, namely the type-II seesaw construction \cite{Mohapatra:1980yp, Lazarides:1980nt, Wetterich:1981bx, Schechter:1981cv, Brahmachari:1997cq} also provides an equally lucrative resolution by introducing a $SU(2)_L$ scalar triplet to the SM field content whose tiny vacuum expectation value (vev) takes care of the small 
neutrino mass. However, to generate the baryon asymmetry of the Universe via leptogenesis, this minimal 
type-II framework needs to be extended either with another triplet \cite{Ma:1998dx,Senami:2003jn,GonzalezFelipe:2013jkc,Lavignac:2015gpa} or by a singlet right-handed neutrino 
\cite{Hambye:2003ka,Hambye:2005tk,Hambye:2012fh,AristizabalSierra:2014nzr,Mishra:2019sye,Rink:2020uvt}. In the latter possibility, the role of the single RHN is to contribute to CP asymmetry generation via the vertex correction (in the triplet decay) provided it carries a Yukawa interaction with the SM Higgs and lepton doublets.  

Additionally, several astrophysical and cosmological observations indicate that energy budget of our Universe requires around 26$\%$ of non-baryonic matter, known as the dark matter (DM) \cite{Julian:1967zz,Tegmark:2003ud,Bennett:2012zja,Clowe:2006eq}. To explain such DM, an extension of the SM is required as otherwise it fails to accommodate any such candidate from its own particle content. Since all these unresolved issues (tiny neutrino mass, matter-antimatter asymmetry and nature of dark matter) point out toward extension(s) of the SM, it is intriguing to establish a common platform for them. With this goal in mind, we focus on the SM extended with a scalar triplet and a fermion singlet (like one RHN). While this can explain the neutrino mass and matter-antimatter asymmetry as stated above, accommodating a DM in it is not that obvious. One simplest possibility emerges if that singlet fermion (the RHN) can be considered as the DM candidate. However, as pointed out above, this fermion taking part in the CP asymmetry generation 
has to carry an Yukawa interaction (of sizable strength) and hence can't be stable provided its mass remains above the electroweak (EW) scale. On the other hand, if it happens to be lighter than the SM Higgs (or gauge bosons), 
it might be a freeze in type of DM \cite{Datta:2021elq}. In this case also, the small Yukawa coupling, as required by the freeze-in generation of DM relic, makes the CP asymmetry negligible and therefore such a possibility needs to be left out. 

We thereby plan to extend this framework by including an inert Higgs doublet (IHD) \cite{LopezHonorez:2006gr,Honorez:2010re,Belyaev:2016lok,Choubey:2017hsq,LopezHonorez:2010tb,Ilnicka:2015jba,Arhrib:2013ela,Cao:2007rm,Lundstrom:2008ai,Gustafsson:2012aj,Borah:2017dfn,Kalinowski:2018ylg,Bhardwaj:2019mts,Borah:2019aeq,Bhattacharya:2019fgs,Bhattacharya:2019tqq,Chakrabarty:2021kmr} such that its lightest neutral component results in 
dark matter while the IHD too contributes to CP asymmetry generation via its Yukawa interaction involving the SM lepton 
doublet and the sole RHN. Involvement of the DM in generating the CP asymmetry required for explaining the matter-antimatter asymmetry of the Universe is an important aspect of our work. Note that the inert doublet DM phenomenology is mostly governed by the gauge interactions and the mass-splitting among the inert Higgs doublet components, but not on the Yukawa interaction \cite{Borah:2017dfn} (contrary to the case of freeze-in RHN as DM) and hence it is not expected to be in conflict with sufficient production of CP asymmetry. Furthermore, search for doubly and singly charged particles involved in the triplet can be quite interesting from collider aspects. Keeping that in mind, we plan to keep the mass of the triplet not very heavy. It is further supported by the finding that the mass splitting among the IHD components (for DM relic satisfaction) along with the neutrino mass generation dominantly by the type-II mechanism keeps the triplet mass below $10^{12}$ GeV. Note that it becomes essential to incorporate the flavor effects in leptogenesis \cite{Abada:2006fw,Nardi:2006fx,Blanchet:2006be,Dev:2017trv,AristizabalSierra:2014nzr,Datta:2021elq,Datta:2021zzf} which come in to effect below the mass equivalent temperature 
$\sim 10^{12}$ GeV. This observation, importance of including flavor effects in triplet leptogenesis, is another salient feature of 
our analysis.

The paper is organized in the following manner. We introduce the structure of the model in Section \ref{sec2} where the particle content with their respective charges under different symmetry group have been discussed. In Section \ref{sec3}, we discuss the mechanism to generate the neutrino mass and how to get a complex structure of the Yukawa coupling matrix responsible for generating the matter-antimatter asymmetry. In section \ref{sec4}, we briefly summarize the inert doublet DM phenomenology and move on to discuss generation of matter-antimatter asymmetry in the Universe via flavor leptogenesis in Section \ref{sec5}. Finally 
in Section \ref{sec6}, we conclude.

\section{The Model} \label{sec2}

The SM is extended with a $SU(2)_L$ scalar triplet $\Delta$, a scalar doublet $\Phi$ and a fermionic SM singlet field $N_R$. The corresponding charge assignments of the relevant fields are provided in Table~\ref{charges}. The Lagrangian involving the new 
fields is then given by
\begin{equation}
-\mathcal{L}_{new}=Y_{\alpha }\bar{\ell}_{L_{\alpha}}\tilde{\Phi}~N_{R}+Y_{\Delta \alpha\beta}\ell^{T}_{L\alpha}Ci\tau_2{\Delta} \ell_{L\beta}+\frac{1}{2}M_{N}\bar{N^{c}_{R}}N_{R}+h.c.,
\label{lag}
\end{equation} 
where $\alpha, \beta$ correspond to three flavor indices. 
Note that $N_{R}$ and $\Phi$ are odd under an additional discrete symmetry $Z_{2}$, thereby making $\Phi$ as inert. This also forbids the Yukawa coupling of the SM Higgs with the RHN, however allows similar interaction with the inert Higgs doublet $\Phi$. The lightest neutral component of this $\Phi$ field plays the role of the dark matter while decay of the triplet into lepton doublets generates the lepton asymmetry which will further be converted into baryon asymmetry by the sphaleron process. Here 
both the inert Higgs doublet and the RHN take part in producing the CP asymmetry. 
\begin{table}[H]
\begin{center}
\begin{tabular}{|r|c|c|c|c|c|c|c|}
\hline
        & $\ell_{L}$ &$e_{R}$& $H$& $N_R$ & $\Delta$ & $\Phi$    \\
\hline        
$SU(2)_L$ &  2  &  1     &  2       &  1     &  3 & 2 \\
\hline
$U(1)_{Y}$&  $-\frac{1}{2}$&$-1$&$\frac{1}{2}$&$0$&$1$ & $\frac{1}{2}$\\
\hline
$Z_2$   & $+$   &  $+$     &   $+$      &  $-$     &   $+$ & $-$ \\
\hline
\end{tabular}
\caption{Particles and their charges under different symmetries.}
\label{charges}
\end{center}
\end{table}

The scalar sector of our model consists of the interaction involving the inert Higgs doublet $\Phi$, Higgs triplet $\Delta$ 
and the SM Higgs $H$. The most general scalar potential for the present scenario can be written as:

\bea
V(H,\D,\Phi)=V_H+V_{\D}+V_{\Phi}+V_{\text{int}},
\eea
where
\besub
\bea
V_H&=&\mu_{H}^{2}(H^{\dagger}H)+\lambda_{H}(H^{\dagger}H)^{2},\\
V_{\D}&=& M^{2}_{\Delta}\mathbf{Tr}(\Delta^{\dagger}\Delta)+\lambda_{\D 1}\mathbf{Tr}(\Delta^{\dagger}\Delta)^2+\lambda_{\D 2}[\mathbf{Tr}(\Delta^{\dagger}\Delta)]^2,\\
V_{\Phi}&=& \mu_{\Phi}^2(\Phi^{\dagger}\Phi)+\lambda_{\Phi}(\Phi^{\dagger}\Phi)^{2},\\
V_{\text{int}}&=& -\mu_{1}(H^{T}i\tau_{2}\Delta^{\dagger} H+h.c)+\lambda_{1}H^{\dagger}H \mathbf{Tr}(\Delta^{\dagger}\Delta)+\lambda_{2}H^{\dagger}\Delta \Delta^{\dagger}H+\lambda_{3}H^{\dagger}\Delta^{\dagger}\Delta H\nonumber\\&+&\lambda_{4}(H^{\dagger}H)(\Phi^{\dagger}\Phi)+\lambda_{5}(H^{\dagger}\Phi)(\Phi^{\dagger}H)+[\frac{\lambda_6}{2}(H^{\dagger}\Phi)^2+h.c]-\mu_{2}(\Phi^{T}i\tau_{2}\Delta^{\dagger} \Phi+h.c)\nonumber\\&+&\lambda_{7}\Phi^{\dagger}\Phi \mathbf{Tr}(\Delta^{\dagger}\Delta)+\lambda_{8}\Phi^{\dagger}\Delta \Delta^{\dagger}\Phi+\lambda_9\Phi^{\dagger}\Delta^{\dagger}\Delta\Phi.
\eea
\eesub
Here we consider all the parameters appearing in the scalar potential to be real. We also consider $\mu_{H}^2<0$ as that would be crucial for electroweak symmetry breaking (EWSB). On the other hand, 
remaining mass parameters such as $\mu_{\Phi}^2, M_{\Delta}^2$ are taken as positive. Denoting the vev of $H$ and $\Delta$ by $v$ (= 246 GeV) and $v_{\Delta}$ respectively, 
the multiplets after the EWSB can be expressed as
\begin{equation}
\Phi =
\begin{bmatrix}
\Phi^{+}\\
\frac{H_{0}+iA_{0}}{\sqrt{2}}
\end{bmatrix}
,H =
\begin{bmatrix}
0\\
\frac{v+h}{\sqrt{2}}
\end{bmatrix}
,\Delta =
\begin{bmatrix}
\frac{\D^{+}}{\sqrt{2}}& \D^{++}\\
v_{\Delta}+\D^0 & -\frac{\D^{+}}{\sqrt{2}}
\end{bmatrix},
\end{equation}
where $h$ is the SM physical Higgs boson with mass 125.09 GeV \cite{deFlorian:2016spz} and 
the induced vev of the triplet is found to be related by \cite{Hambye:2003ka} 
\begin{equation}
v_{\Delta}\simeq \frac{v^{2}\mu_{1}}{2M_{\Delta}^{2}},
\label{delta-vev}
\end{equation}
considering $M_{\Delta} \gg v$. 
Interestingly, the constraint on $\rho$-parameter ($\rho=1.00038\pm 0.00020$)  \cite{ParticleDataGroup:2020ssz} restricts $v_{\Delta} \lesssim$ 4.8 GeV. Note that $v_{\Delta}$ needs to be small enough to accommodate 
the tiny neutrino mass via $\Delta \ell_L \ell_L$ coupling and hence we fix it at 1 eV. Then depending on the mass of 
the $\Delta$ particle, $\mu_1$ can be obtained by the use of Eq.~(\ref{delta-vev}). On the contrary, the analogous coupling $\mu_2$ remains unrestricted and hence can have a sizable value. This therefore will be treated as independent parameter for generating 
sufficient CP asymmetry as we see in the leptogenesis section.

The masses of the different physical scalars of IHD are given (unaffected by the presence of the triplet scalar) as
\bea
m^{2}_{\Phi^{\pm}}&=&\mu_{\Phi}^{2}+\lambda_{1}\frac{v^{2}}{2}, \nonumber \\
m_{H_0}^{2}&=&\mu_{\Phi}^{2}+(\lambda_{4}+\lambda_{5}+\lambda_{6})\frac{v^{2}}{2}, \nonumber \\
m_{A_0}^{2}&=&\mu_{\Phi}^{2}+(\lambda_{4}+\lambda_{5}-\lambda_{6})\frac{v^{2}}{2}.
\label{IHD_masses}   
\eea

%\begin{itemize}
%\item Mass of $A_0$: $m_{A_0}^{2} = m_2^{2}+\frac{\lambda_N}{2}v^2$,
%\item Mass of $H_0$(DM): $m_{H_0}^{2} = m_2^{2}+\frac{\lambda_L}{2}v^2$,
%\item Mass of $\Phi^{\pm}$: $M_{\Phi^{\pm}}^{2} = m_2^2+\lambda_{3}v^2$.
%\end{itemize}
\noindent with $\lambda_L = \frac{\lambda_4+\lambda_5+\lambda_6}{2} >0.$ Without any loss of generality, we consider $\lambda_6 < 0$, $\lambda_5+\lambda_6 < 0$ so that the CP even scalar ($H_0$) is the lightest $Z_2$ odd particle and hence the stable dark matter candidate. Due to the presence of the term proportional to $\mu_1$, there will be a mixing between the SM Higgs and the triplet. However, the mixing being of order $v_{\Delta}$ (taken to be $\sim$ 1 eV, responsible to generate light neutrino mass), this can safely be ignored. We set $\l_1,\l_2,\l_3=0$ for simplicity and then find masses of the physical triplet components as $M_{\D^{\pm\pm}}\simeq M_{\D^{\pm}}\simeq M_{\D^{0}} \simeq M_{\D} $. One should note that LHC puts a strong constraint on mass of $\D^{\pm\pm}$ as $M_{\D^{\pm\pm}}> 820~\text{GeV} ~(870~\text{GeV})$ at $95\%$ C.L. from CMS \cite{Sirunyan:2018koj} (ATLAS \cite{Aaboud:2018xdt}) for $v_{\D}\lesssim 10^{-4}~\text{GeV}$. LHC also set a constraints on $M_{\D^\pm}>350~\text{GeV}$.

%{\color{red}Here we have assumed that the mixing between the Higgs triplet and the SM Higgs is very small, hence it can be neglected and we can choose the coupling $\lambda_{6}$, $\lambda_{7}$, $\lambda_{8}$ to be zero.} 

\section{Neutrino Mass} \label{sec3}

We now proceed to discuss the neutrino mass generation in the present model. As mentioned before, the neutrino mass is expected to be generated via the triplet interaction with the SM lepton doublets resulting the type-II contribution as 
\begin{equation}
m^{\rm{II}}_{\nu} = 2Y_{\Delta} v_{\Delta}.
\label{t-II_relation}
\end{equation}
With a choice of $v_{\Delta}$ as 1 eV, the coupling matrix $Y_{\Delta}$ can be accordingly adjusted to produce the light neutrino matrix ($m_{\nu}$) consistent with the oscillation data. To make it more specific, we consider, 
\begin{align}
	m_{\nu} = m_{\nu}^{\rm{II}} =U^{*} m_{\nu}^{d} U^{\dagger},
	\label{eq:t2nm}
\end{align}
with $m_{\nu}^{d}={\rm{diag}}(m_1,m_2,m_3)$ and $U$ is the PMNS mixing matrix (in the charged lepton diagonal basis) of the form:
\begin{equation}
U=\left(
\begin{array}{ccc}
 c_{12} c_{13} & c_{13} s_{12} & e^{-i \delta } s_{13} \\
 -c_{23} s_{12}-e^{i \delta } c_{12} s_{13} s_{23} & c_{12} c_{23}-e^{i \delta } s_{12} s_{13} s_{23} & c_{13} s_{23} \\
 s_{12} s_{23}-e^{i \delta } c_{12} c_{23} s_{13} & -e^{i \delta } c_{23} s_{12} s_{13}-c_{12} s_{23} & c_{13} c_{23} \\
\end{array}
\right)\times {\rm{diag}}(e^{i\alpha_{1}/2},e^{i\alpha_{2}/2},1),
\end{equation}
parametrized by three mixing angles $\theta_{12},\theta_{23},\theta_{13}$ (denoted by $c_{ij}=\cos\theta_{ij},s_{ij}=\sin\theta_{ij}$), the Dirac CP phase $\delta$ and Majorana CP phases ($\alpha_{1},\alpha_{2})$.
\begin{table}[H]  
\begin{center}
\begin{tabular}{|r|c|c|c|}
	\hline
	Parameter & best fit  & \hphantom{x} 3$\sigma$ range \hphantom{x}
	\\ \hline
	$\Delta m^2_{21} [10^{-5}$eV$^2$]  &  $7.50$   &  6.94--8.14  \\
	\hline
	$|\Delta m^2_{31}| [10^{-3}$eV$^2$]   &  $2.55$  &  2.47--2.63  \\
	\hline
	$\sin^2\theta_{12} / 10^{-1}$         &  $3.18$   &  2.71--3.69  \\
	\hline
	$\sin^2\theta_{23} / 10^{-1}$         &  $5.74$    &  4.34--6.10  \\
	\hline
	$\sin^2\theta_{13} / 10^{-2}$         &  $2.200$   &  2.000--2.405  \\
	\hline
	$\delta/\pi$                          &  $1.08$    &  0.71--1.99  \\
	\hline
\end{tabular}
\end{center} 
\caption{\label{table2} Neutrino mass and mixing parameters from the global fit \cite{deSalas:2020pgw} for NH.}
\end{table}

For simplicity, we now consider Majorana phases and the lightest neutrino mass to be zero. Thereby, using the best-fit values of the mixing angles and $\delta$ \cite{deSalas:2020pgw} as in Table~\ref{table2}, we obtain the following structure 
of the coupling matrix (using Eq.~(\ref{t-II_relation})) in case of normal hierarchy (NH) of neutrinos, 
\begin{equation}
Y_{\Delta}=\left(\frac{1eV}{v_{\Delta}}\right) \times10^{-3}\times\\
\left(
\begin{array}{ccc}
1.84 + 0.27 i & -1.31 -0.75 i & -3.76 - 0.64 i \\
 -1.31 - 0.75  i & 15.94 - 0.08 i & 10.90 + 0.0038 i \\
 -3.76 - 0.64 i & 10.90 + 0.0038 i & 12.09 + 0.07 i \\
\end{array}
\right)
\label{Y-delta}
\end{equation}
We will make use of this $Y_{\Delta}$ in the rest of our analysis wherever appropriate. 

\begin{figure}[t]
	$$
	\includegraphics[scale=0.50]{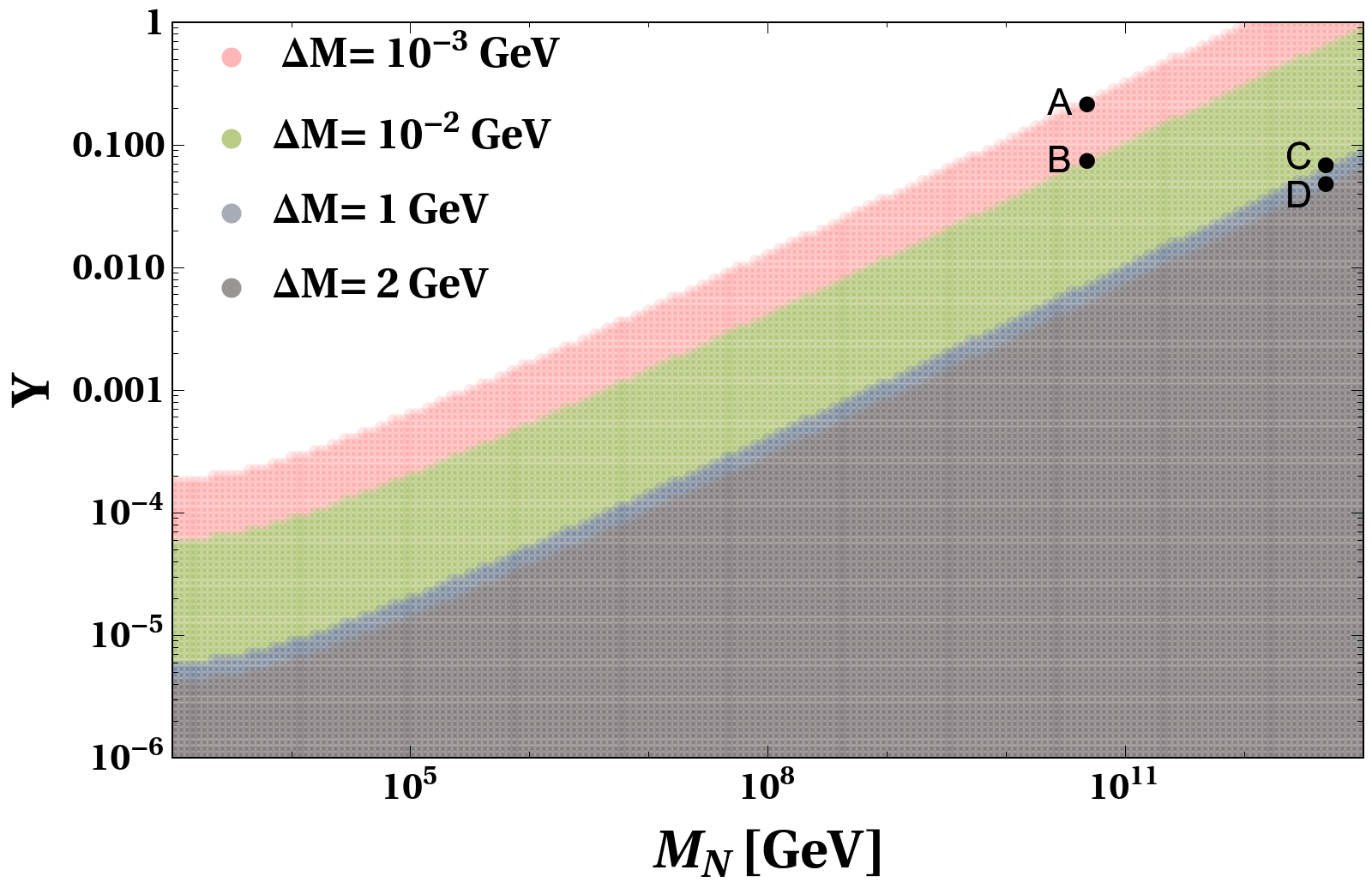}
	$$
	\caption{Allowed range of $Y$ against $M_N$ to keep $m^{\rm{R}}_{\nu}$ subdominant compared to $m^{\rm{II}}_{\nu}$.}
	\label{Y-type-I}
\end{figure}

Note that in our model, due to the presence of one RHN having Yukawa coupling $Y$, a radiative contribution to the light neutrino mass \cite{Ahriche:2017iar} is expected to be present which is given by
\begin{align}
		(m_{\nu}^{\rm{R}})_{\alpha \beta}= \frac{Y_{\alpha}Y_{\beta} M_{N}}{32 \pi^2} \left[\frac{m_{H_0}^2}{m_{H_0}^2-M_{N}^2}\ln\frac{m_{H_0}^2}{M_{N}^2} -\frac{m_{A_0}^2}{m_{A_0}^2-M_{N}^2}\ln\frac{m_{A_0}^2}{M_{N}^2}\right]. 
\end{align}\label{eq:scoto}
%{\color{blue}
%\begin{align}
%	(m_{\nu}^{\rm{R}})_{\alpha \beta}= \frac{Y_{\alpha}Y_{\beta} }{32 \pi^2} \frac{(m_{A_0}+ m_{H_0})\Delta M}{M_N}\left[1+\ln\left(\frac{m_{A_0}^2+m_{H_0}^2}{2 M_N^2}\right)\right]
%\end{align}
%where it is assumed that $M_N^2>>\frac{m_{A_0}^2+m_{H_0}^2}{2}>\frac{m_{A_0}^2-m_{H_0}^2}{2}$.
%}
It is then understood that for a specific value of the DM mass, $m_{H_0}$, along with the mass splitting $\Delta M = m_{\Phi^{\pm}} - m_{H_0} = m_{A_0} - m_{H_0}$, and mass of the RHN, there is a contribution\footnote{Using the fact that $M_N$ is very heavy compared to all IHD components, the radiative contribution can be approximated by $(m_{\nu}^R)_{\alpha \beta} \simeq \frac{Y_{\alpha}Y_{\beta}}{32 \pi^2} \frac{(m_{H_0} + m_{A_0})}{M_N} \Delta M [1+\ln(\frac{m_{H_0}^2 + m_{A_0}^2}{2M_{N}^2})]$.} to the light neutrino mass matrix which depends on the magnitude of $Y_{\alpha}$ coupling. Since we plan to investigate the scenario where the light neutrino mass is mainly contributed by the type-II contribution, we determine here the limits on $Y_{\alpha}$ for which $m^{\rm{R}}_{\nu}$ remains insignificant.

For this purpose, first we assume all $Y_{\alpha}$ to be same given by $Y$. Secondly, we impose a restriction that the contribution to $m_{2}$ (as $m_2$ is the second lightest eigenvalue of $m_{\nu}$) coming from $m^{\rm{R}}_{\nu}$ remains below 10$\%$ contribution followed from type-II seesaw estimate $m^{\rm{II}}_{\nu}$ (henceforth called type-II dominance). Using this ansatz, we provide $Y$ versus $M_N$ plot in Fig.~\ref{Y-type-I} indicating an upper limit on $Y$ value corresponding to a specific RHN mass. In making this plot, we consider DM mass $m_{H^0} =$ 535 GeV with different $\Delta M$ indicated by different colors. This limit on $Y$ will be useful in estimating the CP asymmetry. As the lightest neutrino is taken to be massless in type-II contribution, it is clear that with the appropriate $Y$ value (consistent with the Fig.~\ref{Y-type-I} and leptogenesis), $m_1$ will defer from zero value as it obtains a tiny correction from $m^{\rm{R}}_{\nu}$.

\section{Dark Matter Phenomenology} \label{sec4}

The present setup shelters two particles $N_R$ and $\Phi$ non-trivially charged under $Z_2$. Hence, being stable 
either of them can play the role of the DM. The phenomenology of a singlet fermions like $N_R$ as a WIMP DM candidate with renormalisable interactions remain uninteresting as it predicts overabundant relic density due to the lack of their annihilation channels. On the other hand, as is well known, the study of an IHD provides several interesting prospects 
both in DM phenomenology as well as in collider searches and hence here we primarily stick to the IHD as dark matter 
by considering $M_{N} > m_{H_0}$. An unbroken $Z_2$ symmetry in the current scenario guarantees the stability  of the scalar dark matter. Since it is a well studied framework, in this section, we briefly focus on the parts of DM phenomenology relevant for our analysis extended to leptogenesis section.   

\subsection{Relic Density}

The inert Higgs doublet \cite{LopezHonorez:2006gr,Honorez:2010re,Belyaev:2016lok,Choubey:2017hsq,LopezHonorez:2010tb,Ilnicka:2015jba,Arhrib:2013ela,Cao:2007rm,Lundstrom:2008ai,Gustafsson:2012aj,Kalinowski:2018ylg,Bhardwaj:2019mts,Borah:2019aeq} extension of the SM is one of the simplest extension where a scalar  multiplet can accommodate a DM candidate. Before going into the details of the DM phenomenology of IHD, we first briefly discuss the Boltzmann equation required to study the evolution of the DM in the Universe. DM ($H_0$) being a part of a $SU(2)_L$ doublet always remains in thermal equilibrium in early Universe due to its gauge and quartic interactions. The relic density of such a DM can be calculated by solving the Boltzmann equation 
\bea
\frac{dY_{H_0}}{dz'} &=& -\frac{1}{z'^2} \langle \sigma v_{H_0H_0\rightarrow XX} \rangle 
\left (Y_{H_0}^{2}-(Y_{H_0}^{eq})^2\right )\, ,
\eea
\label{BEidm}
where $z'=m_{H_0}/T$ and $Y_{H_0}^{eq}$ denotes equilibrium number density of $H_0$ whereas $\langle \sigma v_{H_0H_0\rightarrow XX} \rangle$ represents the thermally averaged annihilation cross-section \cite{Gondolo:1990dk} of the DM  annihilating into the SM particles denoted by $X$.  
The relic density of the inert scalar $H_0$ is then expressed as  
\bea
\Omega_{H_0}h^2= 2.755 \times 10^8 \left(\frac{m_{H_0}}{{\rm{GeV}}}\right)Y_{H_0}^0,\, 
\label{relicidm}
\eea
with $Y_{H_0}^0$ denoting the asymptotic abundance of the DM particle after freeze out. In order to calculate the relic density and study the DM phenomenology of the IHD dark matter we use the package micrOMEGAs4.3.5 \cite{Barducci:2016pcb}. 

\begin{figure}[htb!]
\centering
\includegraphics[scale=0.4]{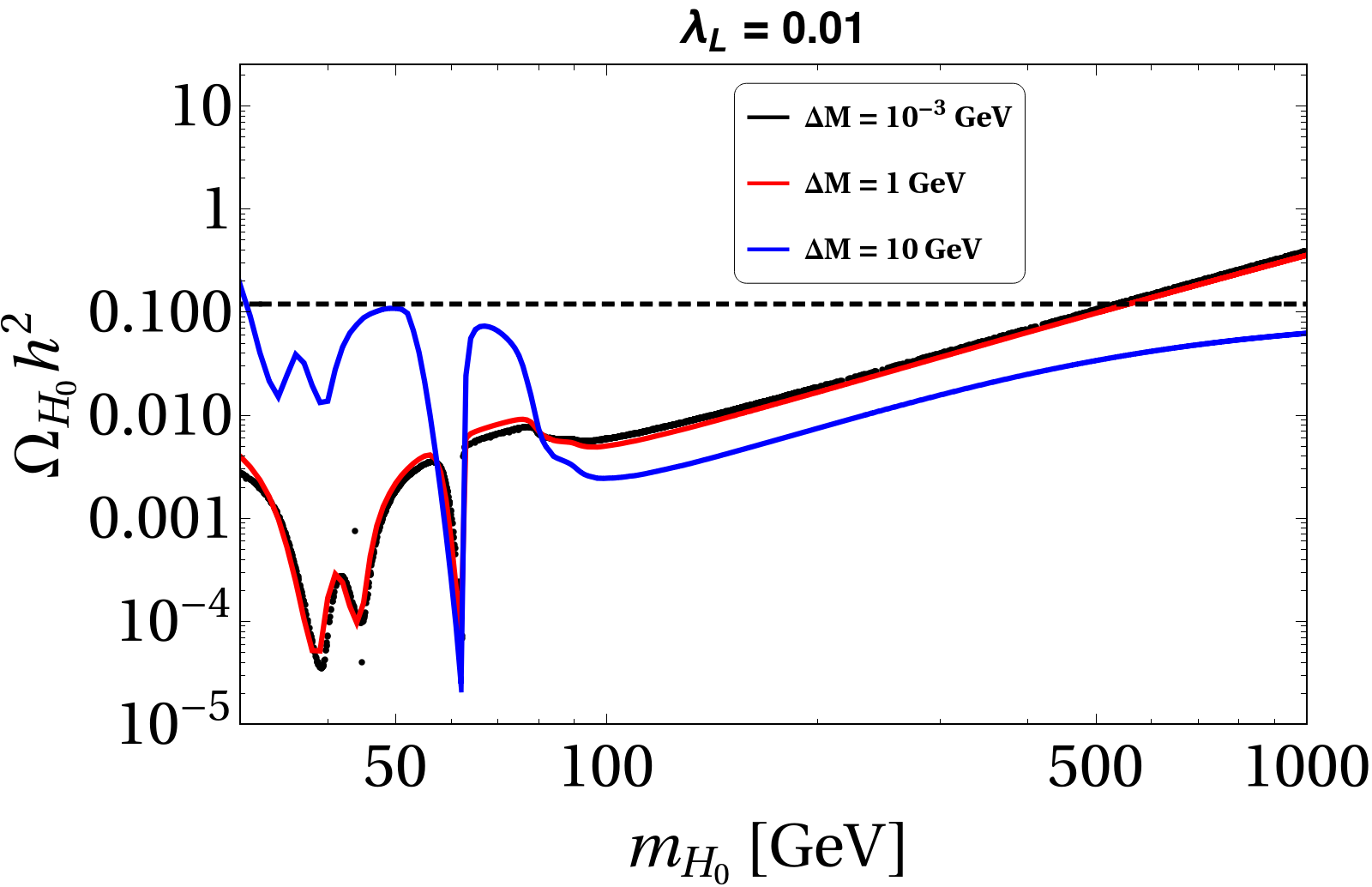}
\caption{Variation of the dark matter relic density ($\Omega_{H_0}h^2$) with its mass ($m_{H_0}$) for
three different values of $\D M$ while keeping $\l_L=0.01$.}
\label{IHD_plot}
\end{figure}

As stated before, the case of an IHD dark matter is well studied and hence, we only summarise the results (in terms of relevant parameters) crucial for our analysis of baryon asymmetry of the Universe in the setup. To facilitate our discussion on DM, we provide a variation of the relic density of $H_0$ ($\Omega_{H_0}h^2$) with the dark matter mass ($m_{H_0}$) for three different choices of mass splitting in Fig.~\ref{IHD_plot}. Here, one notices that in the mass range $M_W\leq m_{H_0} < 500~\text{GeV}$ (popularly known as the intermediate region), a typical feature of the IHD is observed where the relic abundance remains under-abundant. 
 The DM being a part of $SU(2)_L$ doublet, it does annihilate and co-annihilate to the SM gauge bosons with 
a large effective annihilation cross-section resulting this under-abundance in general. However, the correct relic density can still be produced for $m_{H_0}\geq535~\text{GeV}$ with an appropriate choice of mass splitting $\Delta M$ and Higgs portal coupling ($\l_L$) as this leads to the cancellations among 
the $s-$channel, $t-$channel and the contact interaction involved in the scattering amplitude of the DM annihilating into the SM gauge bosons.
In Fig.~\ref{IHD_plot}, one also notices that changing $\Delta M$ from 1 GeV to $10^{-3}$ GeV does not alter the relic result significantly. However, pushing $\Delta M$ to a relatively larger value such as 10 GeV keeps the entire mass range of IHD as under-abundant (except the Higgs resonance region). On the other hand, such a variation in $\D M$ carries significant impact on the DM direct detection experiments as we see in the sub-section \ref{direct_and_indirect} below. Aside such dependence, $\Delta M$ is also found to be restricted from the perturbativity point of view. As shown in \cite{DuttaBanik:2020vfr}, for $\Delta M \gtrsim 20$ GeV, the 
IHD-Higgs coupling along with the Higgs quartic coupling become non-perturbative much before the Planck scale. 

At this stage it is important to point out that the Yukawa interaction of the IHD with the SM leptons and the singlet fermion $N_R$ (see Eq.~(\ref{lag})) does not alter the DM phenomenology of the IHD. Although it introduces an extra annihilation channel for IHD ($t-$channels mediated via $N_R$ of SM leptons), its contribution remains suppressed due to the p-wave suppression. Even though, if this additional annihilation channel has a sizeable contribution to DM abundance compared to other
channels, it does not help generate new allowed DM masses in the intermediate mass regime of the IHD and hence remains uninteresting from the DM point of view.

\subsection{Direct and Indirect Detection}
\label{direct_and_indirect}
The null detection of the DM in direct search experiments like LUX \cite{Akerib:2016vxi}, PandaX-II \cite{Tan:2016zwf, Cui:2017nnn} and Xenon1T \cite{Aprile:2017iyp, Aprile:2018dbl} puts a severe constraints on the DM parameter space. There exists two different possiblities for the DM to interact with nuclei at tree level in the scenario under consideration: (a) elastic scattering 
mediated by SM Higgs boson and (b) inelastic one mediated by electroweak gauge bosons. The spin independent elastic scattering cross section mediated by SM Higgs is given as \cite{Barbieri:2006dq}
\begin{equation}
 \sigma^{\text{SI}} = \frac{\lambda^2_L f^2_n}{4\pi}\frac{\mu_{n}^2 m^2_\mathcal{N}}{m^4_h m^2_{H_0}},
\label{sigma_dd}
\end{equation}
where $\mu_{n} = m_\mathcal{N} m_{H_0}/(m_\mathcal{N}+m_{H_0})$ is the ${\rm DM}$-nucleon reduced mass and $\lambda_{L}$ is the quartic 
coupling involved in ${\rm DM}$-Higgs interaction. 
A recent estimate of the Higgs-nucleon coupling $f$ gives $f = 0.32$ \cite{Giedt:2009mr}. On the other hand, the inelastic scattering cross-section mediated by a gauge boson is expressed as \cite{Cirelli:2009uv},
\bea
\sigma^{\text{SI}}_{\text{IE}}= c\frac{G_F^2m_N^2}{2\pi}Y^2(\mathcal{N}-(1-4s_W^2)\mathcal{Z})^2
\label{inelasticDD}
\eea
 with c = 1 for fermions and c = 4 for scalars. Here the hypercharge of the DM is 1/2. Finally, $\mathcal{N}$ and $\mathcal{Z}$ represents the number of neutrons and protons respectively in the target nucleus with mass $m_{\mathcal{N}}$. With $\D M> 100$ keV \cite{Arina:2009um}, the inelastic scattering of DM with the nuclei is kinematically forbidden as the corresponding cross-section becomes larger than the average kinetic energy of the DM. While $\D M\leq 100$ keV can rule out the entire sub-TeV mass regime of the DM even though allowed by the relic density as $\sigma^{\text{SI}}_{\text{IE}}\simeq4.9\times10^{-8}$ pb which is much larger than the constrained imposed by the Xenon1T experiment in a sub-TeV mass regime of the IHD dark matter.
 
Finally, one should also consider the indirect search experiments like Fermi-LAT \cite{Eiteneuer:2017hoh}, MAGIC \cite{MAGIC:2016xys} etc., which also provide promising detection prospects of the WIMP type DM. These experiments look for an excess of SM particles like photons and neutrinos in the Universe which can be produced from the annihilation or the decay of the DM. The present setup accommodates an IHD dark matter that can also produce such signals which can be detected in the indirect search experiments. The null detection of such signals so far can also constrain the DM parameter space. In a recent study \cite{Borah:2017dfn}, it has been shown that the IHD mass regime below 400 GeV is strictly ruled out by Fermi-LAT.

\section{Leptogenesis} \label{sec5}

\begin{figure}[htb!]
\centering
\includegraphics[scale=0.4]{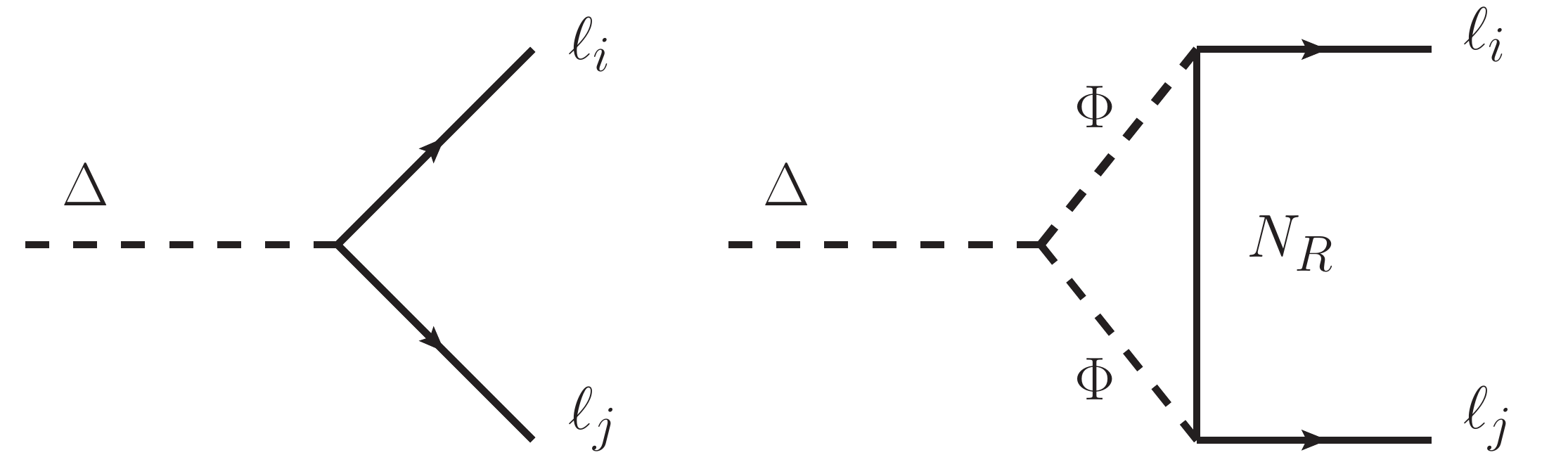}~~
\caption{The tree level and the vertex digram required for the generation of CP asymmetry. }
\label{asy_diagram}
\end{figure}

In this section, we aim to study the leptogenesis scenario resulting from the CP violating out of equilibrium decay of the triplet carrying lepton number of two units in the model. As advocated, this will happen due to the presence of the sole RHN of the setup contributing to the one-loop vertex correction to the tree level triplet decay into leptons as shown in Fig.~\ref{asy_diagram}. It is interesting to note that with one triplet, the generated CP asymmetry can't be of purely flavored one \cite{ AristizabalSierra:2014nzr} in contrast to the presence of this 
possibility in standard triplet leptogenesis involving two scalar triplets. We first discuss the generation of CP asymmetry from the triplet decay and then talk about the evolution of the lepton ($B-L$) asymmetry using Boltzmann equations. In doing so, our plan is to keep the triplet mass as light as possible as that would be interesting from the point of view of collider search for triplet states. In turn, this indicates that flavor effects of the charged lepton Yukawa couplings need to be incorporated provided leptogenesis takes place below temperature $\sim 10^{12}$ GeV.

%\begin{figure}[t]
%	$$	
%	\includegraphics[scale=0.35]{Plots/dp1e-25e10.png}~~
%	\includegraphics[scale=0.35]{Plots/dp2e-25e10.png}~~
%	\includegraphics[scale=0.35]{Plots/dp1e-35e10.png}~~
%	\includegraphics[scale=0.35]{Plots/dp2e-35e10.png}~~
%	$$
%	\caption{$\Delta M=10^{-3}$ GeV, $M_{N}=5 \times10^{10}$ GeV: density plot for $Y$ and $M_{N}$}
%	\label{fig:density-plot}
%\end{figure}
%\begin{figure}[t]
%	$$
%	\includegraphics[scale=0.35]{Plots/dp1e-35e10.png}~~
%	\includegraphics[scale=0.35]{Plots/dp2e-35e10.png}~~
%	$$
%	\caption{$\Delta M=10^{-3}$ GeV, $M_{N}=5 \times10^{10}$ GeV: density plot for $Y$ and $M_{N}$}
%	\label{fig:density-plot}
%\end{figure}
\subsection{CP asymmetry generation}

The flavored CP asymmetry produced as a result of the interference between the tree level and the loop level diagram 
shown in Fig.~\ref{asy_diagram} can be defined and evaluated as~\cite{Hambye:2003ka,Hambye:2005tk}
\begin{align}
 \epsilon_{\Delta}^{\ell_i} & = 2\frac{\sum_{j} \Gamma(\bar{\Delta}\rightarrow \ell_i+\ell_j)-\Gamma(\Delta\rightarrow \bar{\ell}_i+\bar{\ell}_j)}{\Gamma_{\Delta}^{\textrm{tot}}+\Gamma_{\bar{\Delta}}^{\textrm{tot}}},\\
&= \frac{1}{4\pi}M_{N}\frac{\sum_{j} \mathbf{Im}\left[\mu_{2} Y_{i}Y_{j }(Y_{\Delta})_{ij}\right]}{\mathbf{Tr}(Y_{\Delta}^{\dagger} Y_{\Delta})M_{\Delta}^{2}+|\mu_1|^2+|\mu_2|^2}\log\left(1+\frac{M_{\Delta}^2}{M_{N}^2}\right),
 \label{eq:cp}
 \end{align}
where, $\Gamma_{\Delta}^{\textrm{tot}}$ is the total decay width of $\Delta$:
\begin{align}
	\Gamma_{\Delta}^{\textrm{tot}} &=\sum_{i,j} \Gamma(\Delta \to \bar{\ell}_i \bar{\ell}_j) + \Gamma(\Delta \to H H) +\Gamma(\Delta \to \Phi \Phi), \\
&=\frac{M_{\Delta}}{8\pi} \left[\mathbf{Tr}(Y_{\Delta}^{\dagger} Y_{\Delta})+ \frac{|\mu_1|^2+|\mu_2|^2}{M_{\Delta}^2}\right].
\end{align}
Similarly, the anti-triplet decay $\Gamma_{\bar{\Delta}}^{\textrm{tot}}$ also contributes to the total decay width in the denominator. It would be useful to define the branching ratios $B_\ell, B_H$ and $B_{\Phi}$ at this stage, representative 
of the $\Delta$ triplet decay to lepton and scalar final states as:
\begin{align}
\label{eq:lepton-and-scalar-BRs}
B_\ell
&=\sum_{i=e,\mu,\tau}B_{\ell_i}
=\sum_{i,j=e,\mu,\tau}B_{\ell_{ij}}
 = \sum_{i,j=e,\mu,\tau}
\frac{M_{\Delta}}{8\pi  \Gamma^\text{Tot}_{\Delta}}
|(Y_{\Delta})_{ij}|^2\ ,
\nonumber\\
B_H &=\frac{|\mu_1|^2}
{8\pi M_{\Delta}\Gamma^\text{Tot}_{\Delta_\alpha}}\ , \qquad B_{\Phi}= \frac{|\mu_2|^2}
{8\pi M_{\Delta}\Gamma^\text{Tot}_{\Delta_\alpha}}\ ; \qquad B_{\ell}+B_H +B_{\Phi} =1
\end{align}

We notice now that among the various parameters involved in the expression of flavored CP asymmetry $\epsilon_{\Delta}^{\ell_i}$ in Eq.~(\ref{eq:cp}), $Y_{\Delta}$ is obtained from Eq.~(\ref{Y-delta}) while $\mu_1$ becomes function of $M_{\Delta}$ 
via Eq.~(\ref{delta-vev}) with the choice $v_{\Delta}$ = 1 eV. Finally, to maximize the CP asymmetry, we fix $Y$ to its largest allowed value corresponding to a specific choice of $M_{N}$ (and $\Delta M$) from Fig.~\ref{Y-type-I} so as to restrict the radiative contribution negligible (keeping it below 10$\%$) compared to the type-II one toward light neutrino mass. Although there is no direct correlation between CP asymmetry and the mass splitting $\Delta M$ among IHD components, it can be noted that $\Delta M$ being involved in restricting the maximum value of $Y$ for the type-II dominance of neutrino mass (see Fig.~\ref{Y-type-I}), plays an important role here.  
Hence, $\epsilon_{\Delta}^{\ell_i}$ effectively remains function of three independent parameters $\mu_2, M_{\Delta}$ and $M_{N}$. It is interesting to note that in this case, there exists a coupling $\mu_2$ in the CP asymmetry expression which does not participate in the neutrino mass generation unlike conventional type-(I+II) scenario where all the couplings involved in CP asymmetry also take part in the neutrino mass \cite{Hambye:2003ka,Hambye:2005tk,Rink:2020uvt}. As a result, in the latter case ({\it{i.e.}} in type-(I+II)) with type-II dominance, the relevant parameter space is restricted leading to $M_{\Delta}$ quite heavy. For example, it was shown in \cite{Rink:2020uvt}, in the context of type-II-dominated left-right seesaw model, that $M_{\Delta}$ turns out to be $10^{12}$ GeV or beyond. On the other hand, involvement of otherwise free parameter $\mu_2$ may open up a relatively wider parameter space in our case. Below we proceed to get some idea on the CP asymmetry generation by scanning over the parameters for our work. 

\begin{figure}[H]
	\centering
	\subfigure[]{\includegraphics[width=60mm]{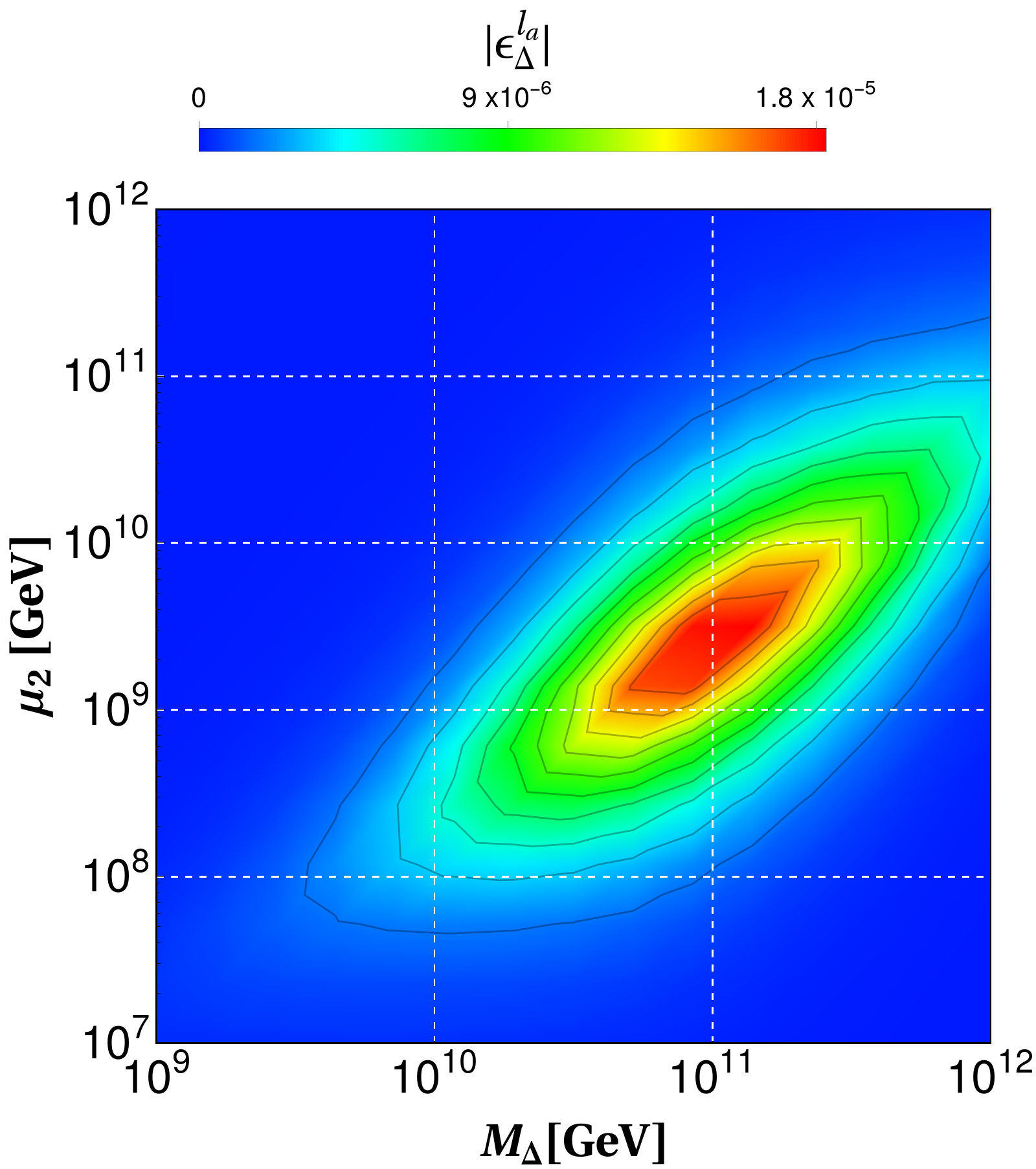}}
	\subfigure[]{\includegraphics[width=60mm]{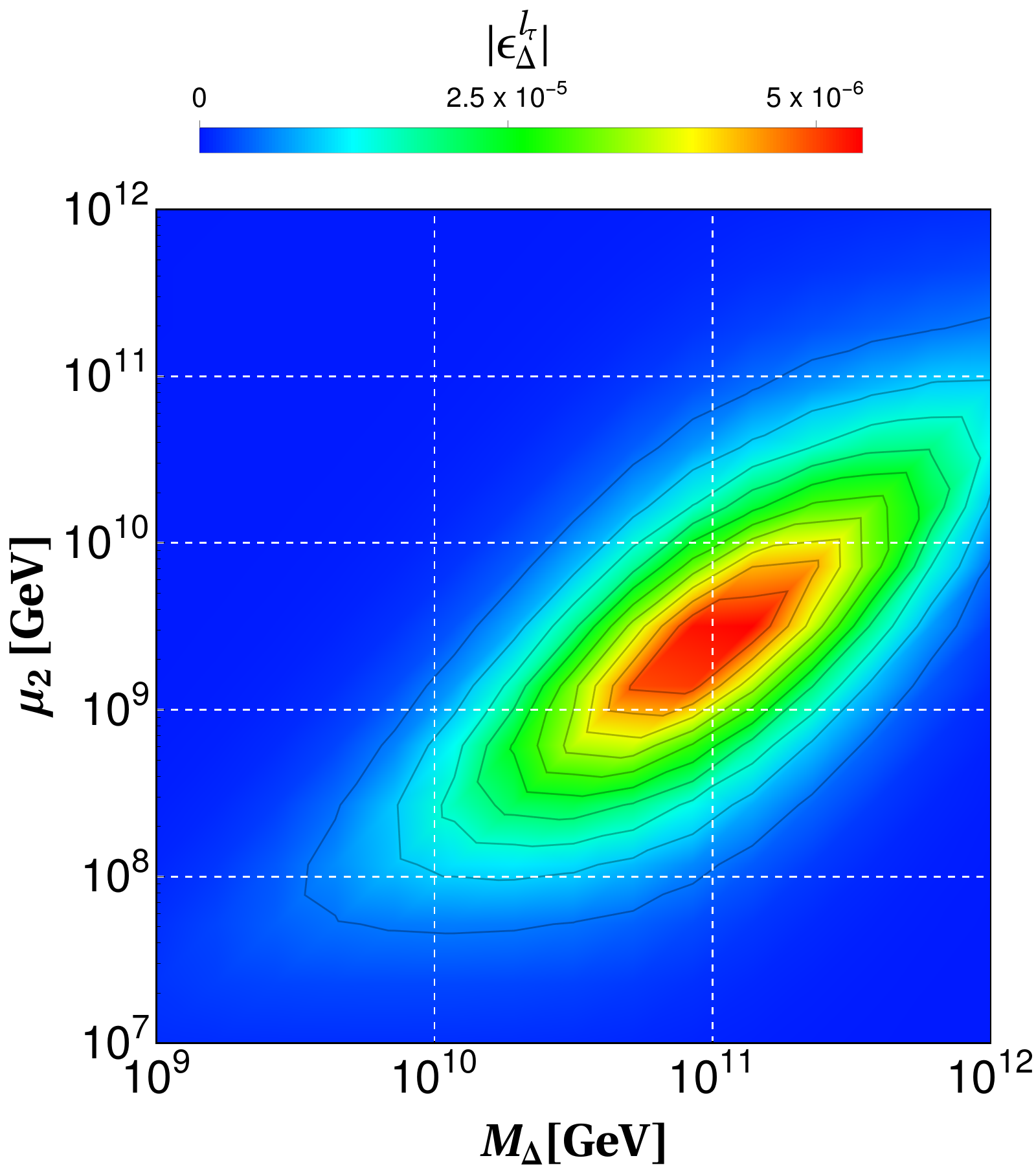}}
	\subfigure[]{\includegraphics[width=60mm]{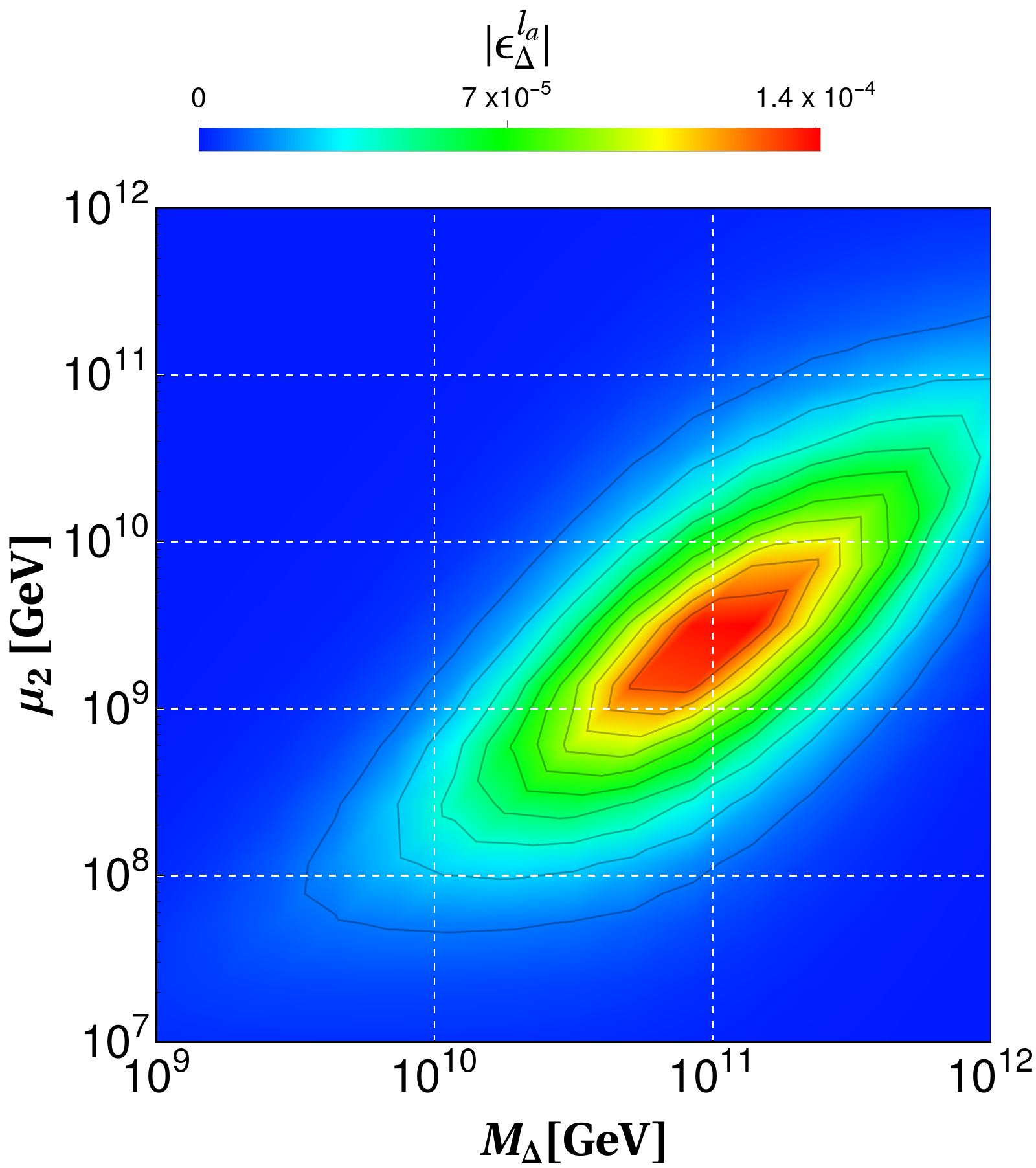}}
	\subfigure[]{\includegraphics[width=60mm]{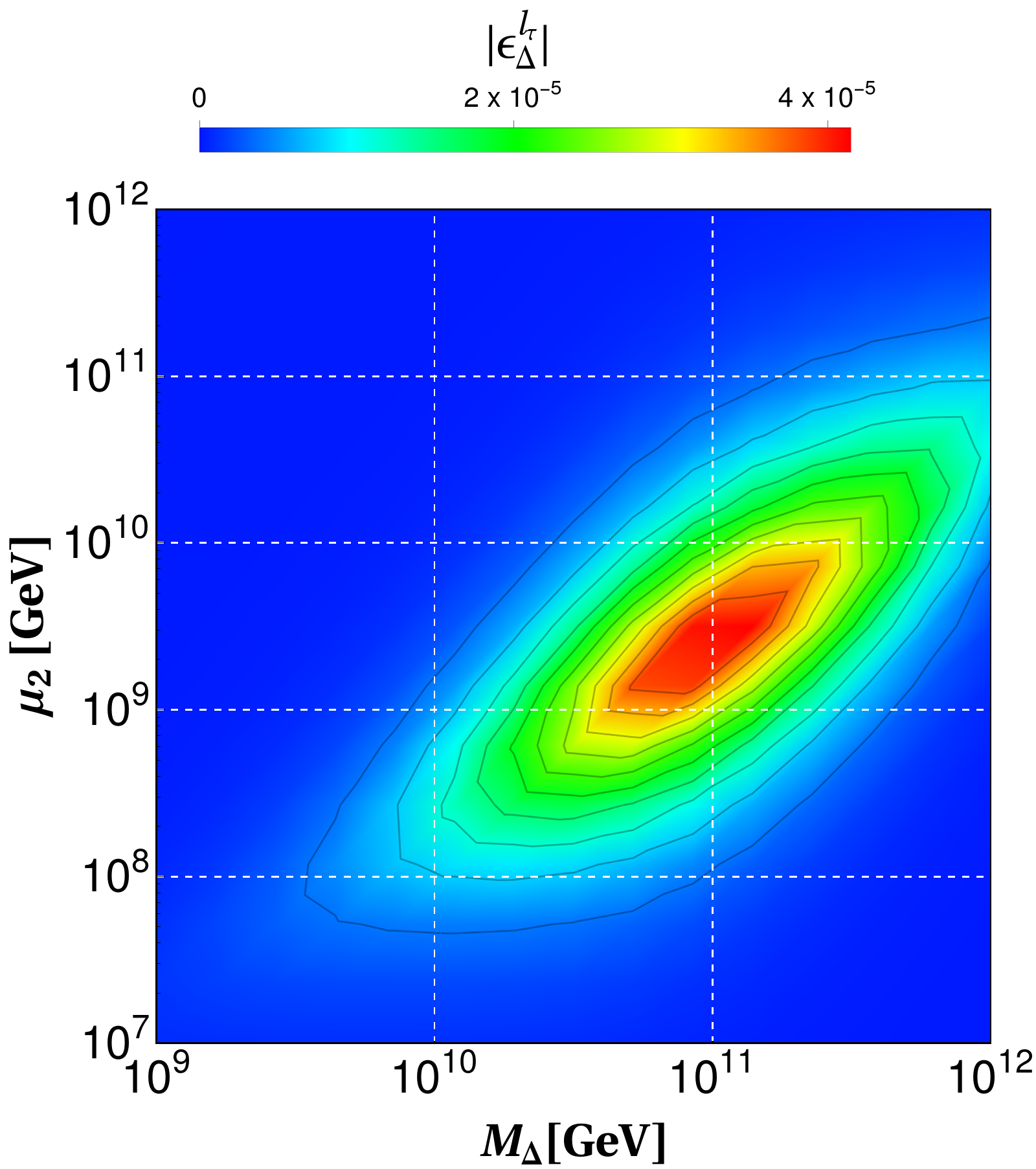}}
	\caption{Density contour of individual CP asymmetry in $\mu_2 - M_\Delta$ plane with $M_N=5\times10^{10}$ GeV. Top panel: $\D M=10^{-3}$ GeV (using BP [A] from Fig.~ \ref{Y-type-I}) ; Bottom panel: $\D M=10^{-2}$ GeV (using BP [B] from Fig.~\ref{Y-type-I}). }
	\label{fig:density-plot}
\end{figure}

Fig.~\ref{fig:density-plot} shows the density contour plot for the absolute value of individual components of CP asymmetry in the $M_\Delta - \mu_2$ plane while keeping $M_{N}$ fixed at a specific value $5 \times 10^{10}$ GeV. In producing these 
plots, we also consider two specific choices of $\Delta M = 10^{-3}$  GeV (top panel) and $\Delta M = 10^{-2}$ GeV (bottom panel).  Benchmark Points (BP) [A] and [B] are used to specify the values of Yukawa coupling $Y$ for these mass splittings respectively (see Fig.~\ref{Y-type-I}).
For $M_{\Delta}$ below $10^{11}$ GeV (though larger than $10^9$ GeV), tau-Yukawa interaction comes to equilibrium making the asymmetries along $a$ 
(a coherent superposition of $e$ and $\mu$ lepton flavors) and $\tau$ flavor distinguishable. We elaborate on it latter. 
Hence in this region, we study $\epsilon^{\ell_{a, \tau}}_{\Delta}$ separately in Fig.~\ref{fig:density-plot}(a) and (b) respectively. In the plot, the red (blue) region indicates the highest (lowest) absolute value of CP asymmetry while the green and the yellow regions indicate intermediate values of it. Maximum CP asymmetry\footnote{For the present analysis, it turns out $B_l >> B_{\Phi}$.} (value of which is indicated in the top bar) results around the central region. Comparing the top and bottom panels of plots, we find that it is possible to generate relatively larger CP asymmetry once we lower $\Delta M$. However, we can not keep on lowering $\Delta M$ indefinitely ($\Delta M$ below 100 keV is not feasible as we have discussed it 
in the DM section). 

\subsection{Evolution of $B-L$ asymmetry}
With the above estimates of the CP asymmetries in different flavor directions, we study here the evolution of $B-L$ asymmetry via Boltzmann equations. It naturally involves all possible interactions with the thermal bath in the early 
Universe. As we aim here to bring down the leptogenesis scale (i.e. as low $M_{\Delta}$ as possible) as stated earlier, 
the situation becomes more involved. Below the energy scale $M_{\Delta} \lesssim 10^{12}$ GeV, particularly in the temperature regime $10^{9}$ GeV $\lesssim T\lesssim 10^{12}$ GeV, along with bottom and charm quarks, tau Yukawa related interactions come to thermal equilibrium because of which the quantum coherence of lepton doublets is lost. As a result, in this regime, two orthogonal directions denoted by $i = a$ (coherent superposition of $e$ and $\mu$ lepton flavors) and $\tau$ survive. 
On the other hand, QCD instanton and EW sphaleron reactions also reach equilibrium at this temperature range making the baryon number as a non-conserved quantity, though it conserves the individual $B/3-L_i$ charges. So, an appropriate study of the evolution of the lepton asymmetry should be performed by knowing the evolution of the $B/3 - L_i$ charges with $i = a$ and $\tau$. Further below region of $10^5$ GeV $\lesssim T\lesssim 10^9$ GeV, strange quark and muon Yukawa interactions achieve thermal equilibrium indicating that the lepton doublets completely lose their quantum coherence. Hence, lepton asymmetry becomes distinguishable along all three flavors $e,\mu$ and $\tau$. Below $T<10^5$ GeV, electron Yukawa reaches the equilibrium.

In order to study the evolution of the $B-L$ asymmetry, we need to employ a set of coupled Boltzmann equations following the analysis of \cite{AristizabalSierra:2014nzr}. This set includes differential equations for triplet density $\Sigma= \Delta + \bar{\Delta}$, triplet asymmetry $\Delta_{\Delta}=\Delta - \bar{\Delta}$ and $B/3- L_i$ asymmetries considering flavor effects  as we have considered a specific hierarchy $M_{\Delta} < M_{N}$ in our analysis. Assuming the triplet scalar was at thermal equilibrium with plasma in the early Universe, below are the specified interactions which 
have the potential to change its number density as well as produce or washout the effective $B-L$ charge asymmetry:
\begin{itemize}
	\item Decay [$\Delta \rightarrow \bar{\ell}_i\bar{\ell}_j$, $\Delta \rightarrow H H$ and $\Delta \rightarrow \Phi \Phi$] and inverse decay: The
	total decay rate density is then represented by:  
	$\gamma_{D}=\gamma_{D}^\ell + \gamma_{D}^H+\gamma_{D}^\Phi$, with 
	$\gamma_D=\frac{K_1(z)}{K_2(z)}\,n^\text{Eq}_\Sigma\,\Gamma^\text{Tot}_\Delta \ ,$ $K_1(z),K_2(z)$ are the modifed Bessel functions. Here $n^\text{Eq}_\Sigma$ is the equilibrium number desity of $\Sigma_{\Delta}$ and $z$ is defined as $M_{\Delta}/T$.
	\item Gauge induced scatterings $\Delta\Delta \leftrightarrow f~f$, $\Delta\Delta \leftrightarrow XX$ (s-channel), $\Delta\Delta \leftrightarrow GG$ (triplet mediated t,u-channel and four point vertex contrubutions), where $f$ stands for SM fermions, $G$ are SM Gauge bosons, $X=H,\Phi$. Altogether the reaction densities are characterized by $\gamma_{A}$, where 
	$\gamma_A=\frac{m_\Delta^4}{64\,\pi^4}\int_{x_\text{min}}^\infty
	\,dx\sqrt{x}\,\frac{K_1(z\sqrt{x})\;\widehat\sigma_A}{z}\ ,$ $x=s/M_{\Delta}^2$ (s is the centre of mass energy). Here $\widehat\sigma_A$ is the reduced cross section inclusive of all gauge induced processes, where 
	$\widehat\sigma$ is related to the usual cross section $\sigma$ for a process $[1+2\to 3+4+...]$ by:
	$\widehat\sigma= \frac{8}{s}[(p_{1}.p_{2})^2- m_1^2 m_2^2] \sigma$ with $p_1,p_2$ be the four momentum of initial particles having mass $m_1,m_2$.
	
	\item Lepton number ($\Delta L = 2$) and Lepton flavor violating $s$ and $t$ channel scatterings (mediated by 
	the triplet/anti-triplet):
	$X X \leftrightarrow  \bar{\ell}_i\bar{\ell}_j$, $X \ell_j \leftrightarrow \bar{X}\bar{\ell}_i$ having reaction densities 
	$\gamma^{XX}_{\ell_i\ell_j}$ and $\gamma^{X\ell_j}_{X \ell_i}$ respectively. 
	
	\item Lepton flavor violating triplet mediated s and t channel scattering: 
	$(\ell_a \ell_b \leftrightarrow \ell_i \ell_j)_s$, $(\ell_a \ell_b \leftrightarrow \ell_i \ell_j)_t$ with reaction densities given by
	$(\gamma^{\ell_a\ell_b}_{\ell_i\ell_j})_s$ and
	$(\gamma^{\ell_a\ell_b}_{\ell_i\ell_j})_t$.
\end{itemize}

Keeping in mind the above discussion, the following Boltzmann equations are constructed, 
\begin{align}
\label{eq:flavored-BEqs1}
% -------------- Equation 1 --------------
s H z\frac{dY_{\Sigma}}{dz}&=
-\left(
\frac{Y_{\Sigma}}{Y^\text{Eq}_\Sigma} - 1
\right)\gamma_{D}
-2\left[
\left(\frac{Y_{\Sigma}}{Y^\text{Eq}_\Sigma}\right)^2 - 1
\right]\gamma_{A}\ ,
\\
\label{eq:flavored-BEqs2}
% ------------ Equation 2 ----------------
s H z\frac{dY_{\Delta_{\Delta}}}{dz}&=
-\left[
\frac{Y_{\Delta_{\Delta}}}{Y^\text{Eq}_\Sigma}
-
\sum_k
\left(
\sum_i B_{\ell_i}C^\ell_{ik}
-
B_H C^H_k
\right)\frac{Y_{X_k}}{Y^\text{Eq}_\ell}
\right]\gamma_{D}\ ,
\\
% ------------- Equation 3 ---------------
\label{eq:flavored-BEqs3}
s H z\frac{dY_{\Delta_{B/3-L_i}}}{dz}&=
-\left(\frac{Y_{\Sigma}}{Y^\text{Eq}_\Sigma}-1\right)
\epsilon^{\ell_i}_{\Delta}
\gamma_{D}
+ 2\sum_j
\left(
\frac{Y_{\Delta_{\Delta}}}{Y^\text{Eq}_\Sigma}
-\frac{1}{2}\sum_k
C^\ell_{ijk}\,\frac{Y_{X_k}}{Y^\text{Eq}_\ell}
\right)B_{\ell_{ij}}\gamma_{D}
\nonumber\\
&-\sum_{j,k}
\left\{2\left(
C^H_k
+
\frac{1}{2} C^\ell_{ijk}
\right)
\left(
\gamma^{HH}_{\ell_i\ell_j}
+
\gamma^{H\ell_j}_{H\ell_i}
\right)+
C^{\ell}_{ijk}
\left(
\gamma^{\Phi\Phi}_{\ell_i\ell_j}
+
\gamma^{\Phi\ell_j}_{\Phi\ell_i}
\right)\right\}
\frac{Y_{X_k}}{Y^\text{Eq}_\ell}
\nonumber\\
&
-\sum_{j,a,b,k}
C^\ell_{ijabk}\,
\left(
(\gamma^{ \ell_a \ell_b}_{\ell_i\ell_j})_s
+
(\gamma^{ \ell_a \ell_b}_{\ell_i\ell_j})_t
\right)
\frac{Y_{X_k}}{Y^\text{Eq}_\ell},
\end{align}
where $Y_{\Delta_X}$ is defined as the ratio between particle and antiparticle number densities difference 
to entropy: $Y_{\Delta_X}= (n_X-n_{\bar{X}})/s$, where $n_X(n_{\bar{X}})$ is number density of a particular species $X(\bar{X})$. 

Depending on the temperature range, the  index $i$ in the RHS of Eq.~(\ref{eq:flavored-BEqs3}) will run differently, $e.g. ~$ for $10^9$ GeV $\lesssim T\lesssim10^{12}$ GeV, $i=a,\tau$ as done in Fig.~\ref{fig:density-plot}, while for $T < 10^9$ GeV, $i=e,\mu,\tau$ need to be included. The generated asymmetry in number densities involving leptons of a specific flavor $Y_{\Delta \ell_i}$ as well as that of the Higgs (originated from the inverse decay and subtraction of the 
on-shell contribution for $\Delta L =2$ processes) can be related to the fundamental asymmetries $\Delta_{\Delta}$ and 
$\Delta_{B/3-L_i}$ with the help of the equilibrium conditions applicable. 
The corresponding conversion factors\footnote{We simplify the situation by considering the chemical potential of the $\Phi$ field to be zero and hence corresponding $C^{\Phi}$ does not appear.} are defined in terms of $C^\ell$ and $C^H$ matrices as below \cite{AristizabalSierra:2014nzr,Nardi:2006fx}:  
\begin{equation}
\label{eq:lepton-doublets-scalar-doublet-BmL-asymm}
Y_{\Delta_{\ell_i}}=-\sum_k \,C^\ell_{ik}\,Y_{X_k}
\qquad
\mbox{and}
\qquad
Y_{\Delta_H}=-\sum_k \,C^H_k\,Y_{X_k},\ 
\end{equation}
where $Y_{X_k}$ are the elements\footnote{Here we take $Y_{\Delta_\Delta}\equiv Y_{\Delta_{\Delta^0}}=Y_{\Delta_{\Delta^+}}=Y_{\Delta_{\Delta^{++}}}$  and $Y_{\Delta_H}\equiv Y_{\Delta_{H^0}}=Y_{\Delta_{H^+}}$.} 
of 
$Y_{X}^T=
\begin{pmatrix}
Y_{\Delta_\Delta},
Y_{\Delta_{B/3-L_k}}
\end{pmatrix}
$ 
and $C^\ell_{ijk}$ and $C^\ell_{ijabk}$ are given by:
\bea
\label{eq:flavor-coupling-matrices}
C^\ell_{ijk}=C^\ell_{ik} + C^\ell_{jk}\ ,~~
C^\ell_{ijabk}=C^\ell_{ik} + C^\ell_{jk} -C^\ell_{ak} - C^\ell_{bk}.
\eea
Then the final lepton asymmetry is converted to Baryon asymmetry via sphaleron processes as given by :
\begin{align}
 Y_{\Delta B}= 3 \times \frac{12}{37} \sum_i Y_{\Delta_{B/3-L_i}}, 
 \label{final-B}
\end{align}
where the factor $3$ is due to the degrees of freedoms associated to the $SU(2)_L$ scalar triplet. 

\subsection{Results}

\begin{figure}[htb!]
	$$
	\includegraphics[scale=0.4]{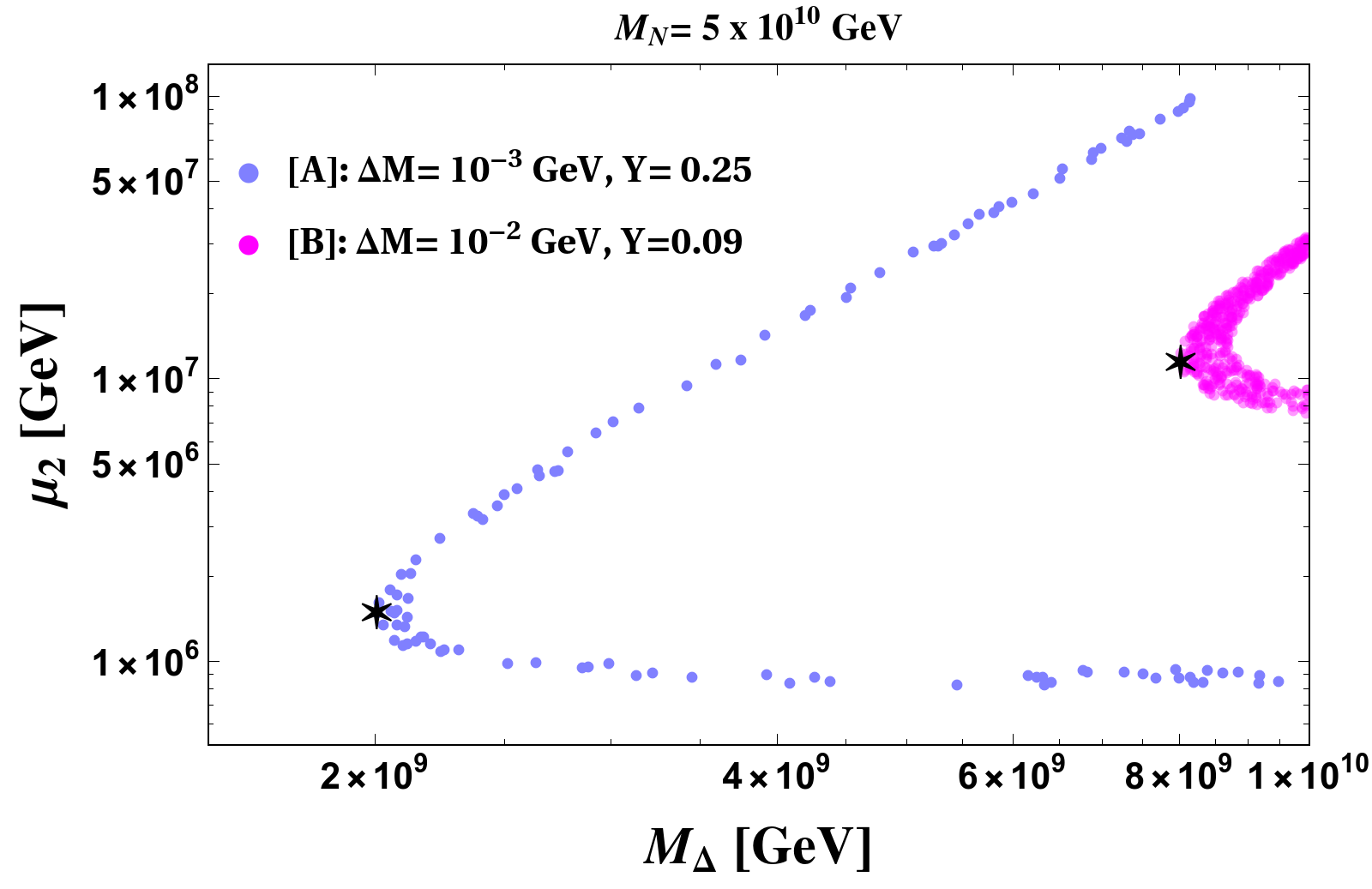}~~
	\includegraphics[scale=0.4]{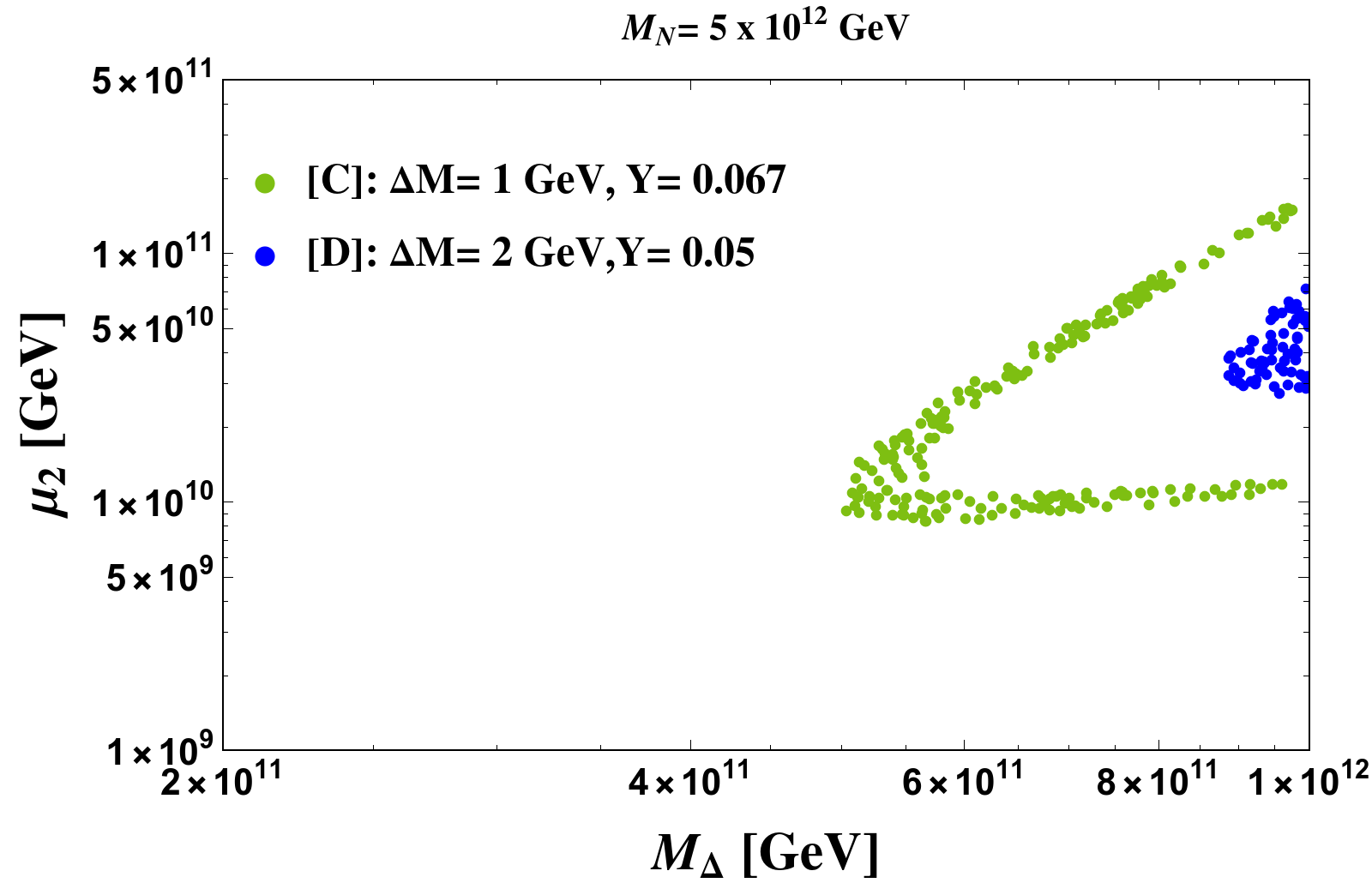}~~
	$$
	\caption{Contour plots for final baryon asymmetry ($Y_{\Delta_B} = (8.718 \pm 0.012)\times 10^{-11}$) in $\mu_2-M_\D$ plane. Left panel: $M_N=5\times10^{10}~\text{GeV}$ while $\D M=10^{-3}$ GeV (light blue) and $\D M=10^{-2}$ GeV (magenta), Right panel: $M_N=5\times10^{12}~\text{GeV}$ while $\D M=1$ GeV (light green) and $\D M=2$ GeV (blue).}
	\label{fig5}
	\end{figure}

In order to explore the parameter space of our model so as to produce the observed baryon asymmetry $Y_{\Delta_B} = (8.718 \pm 0.012)\times 10^{-11}$ \cite{Cyburt:2015mya,Planck:2018vyg}, first we choose a specific value of $M_{N}=5\times 10^{10}$ GeV. Then 
based on our previous discussion we infer that lepton asymmetry with $10^9$ GeV$< M_{\Delta} < M_N$ will be produced along two orthogonal directions, $i.e.$ $a$ and $\tau$ while below $10^9$ GeV,  all three flavor directions have to be taken into account. With the help of chemical equilibrium constraint equations (coming from 
relevant Yukawa and EW Sphaleron related reactions that are in equilibrium) as well as other constraints such as hypercharge conservation (applicable in this energy range) lead to the following structure of $C^{\ell}$ and $C^{H}$ matrices (for $10^9$ GeV$< M_{\Delta} < M_N$) \cite{Nardi:2006fx,AristizabalSierra:2014nzr}:
\begin{align}
\label{eq:C-ell-PFL-used}
C^\ell=\frac{1}{718}
\begin{pmatrix}
- 12 & 307 &- 36 \\
78 &- 21 &  234
\end{pmatrix}, \ 
C^H=\frac{1}{359}
\begin{pmatrix}
258 &  41 & 56
\end{pmatrix}.\ 
\end{align}
For $M_{\Delta}$ below $10^9$ GeV, $C^{\ell}$ and $C^H$ become $3 \times 4$ and $1 \times 4$ matrices.

Using the input on the flavored CP asymmetries along $a$ and $\tau$ directions from Fig.~\ref{fig:density-plot}, obtained as a function of $\mu_2$ and $M_{\Delta}$ for a specific $\Delta M$ value,  we employ the set of Boltzmann equations  Eq.~(\ref{eq:flavored-BEqs1}-\ref{eq:flavored-BEqs3}) while ignoring the last two terms of Eq.~(\ref{eq:flavored-BEqs3}) to draw a contour plot for the correct final baryon asymmetry (via Eq.~(\ref{final-B})) as shown in Fig.~\ref{fig5}. The two contour plots, one in magenta and other in light blue patches, correspond to $\Delta M = 10^{-2}$ GeV and $10^{-3}$ GeV respectively. 
We therefore infer that the triplet can be as light as $\sim 10^9$ GeV, contrary to the type-(I+II) case, which can successfully generate the required amount of baryon asymmetry, thanks 
to the flexibility involved due to the presence of parameter $\mu_2$. 

We also notice that with the increase in $\Delta M$ value, the baryon asymmetry satisfying contour gets shifted toward heavier mass range of the triplet. This observation is interesting as it is correlated to the DM phenomenology. We recall that even though a smaller $\Delta M$ is not in conflict with the relic contribution to DM (in fact a relic satisfaction requires $\Delta M$ below $\mathcal{O}(1)$ GeV), it is actually restricted from below by the inelastic scattering of DM direct searches ($\Delta M \gtrsim 10^{-4}$ GeV). On the other hand, the upper bound on $\D M$ follows from the relic density satisfaction by the DM as can be seen from the Fig.~\ref{IHD_plot}. We find that with $\D M=1~\text{GeV}$ (light green), $M_{\D}$ can be as low as $5\times10^{11}~\text{GeV}$, while a further increase in $\D M$ such as 2 GeV (blue)\footnote{With such mass splitting, the DM having mass $m_{H^0}= 535$ GeV can constitute $\sim83\%$ of the observed relic abundance.} pushes the lightest possible $M_{\D}$ value to $9\times10^{11}~\text{GeV}$ (while maintaining $M_{\D} < M_{N}$) (see Fig.~\ref{fig5}(b)). As a result, we can work in the two flavor regime of leptogenesis with such $\D M$ values. A further increase in $\D M$ will take us to unflavored regime of leptogenesis. However, we refrain from considering  a further larger values of $\D M$ mainly because in that case, 
the relic density of the DM becomes under-abundant.
\begin{figure}[t]
	\label{fig6contour}
	\includegraphics[scale=0.4]{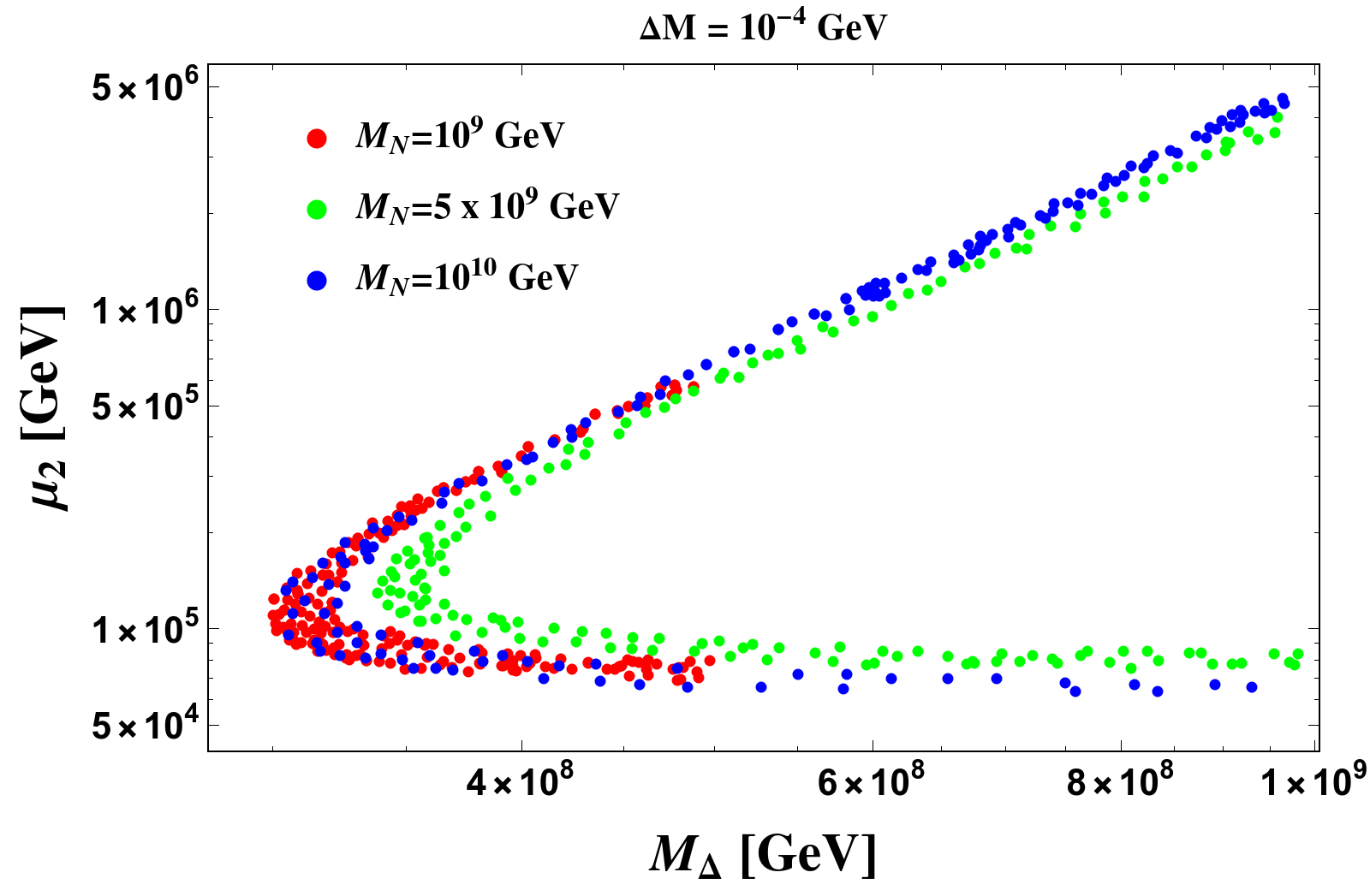}~~
	\caption{Contour plots for final baryon asymmetry ($Y_{\Delta_B} = (8.718 \pm 0.012)\times 10^{-11}$) in $\mu_2-M_\D$ plane for $\D M=10^{-4}$ while $M_N=10^{10}~\text{GeV}$ (blue),  $M_N=5 \times 10^{9}~\text{GeV}$ (green) and $M_N= 10^{9}~\text{GeV}$ (red).}
	\end{figure}
In Fig.~\ref{fig6contour}, a similar contour plots are presented, but for fixed $\Delta M = 10^{-4}$ GeV with different $M_{N}$ values. As can be seen, the allowed mass of the triplet comes down to an even lower mass close to $10^8$ GeV. It is perhaps pertinent here to mention that our entire parameter space corresponds to the Yukawa regime \cite{Hambye:2012fh, AristizabalSierra:2014nzr} 
where the Yukawa induced inverse decay processes (characterised by $B_{\ell_{ij}}\gamma_D$) play important role and hence flavor effects become crucial.

\begin{figure}[htb!]
\centering
\subfigure[]{
	\includegraphics[scale=0.4]{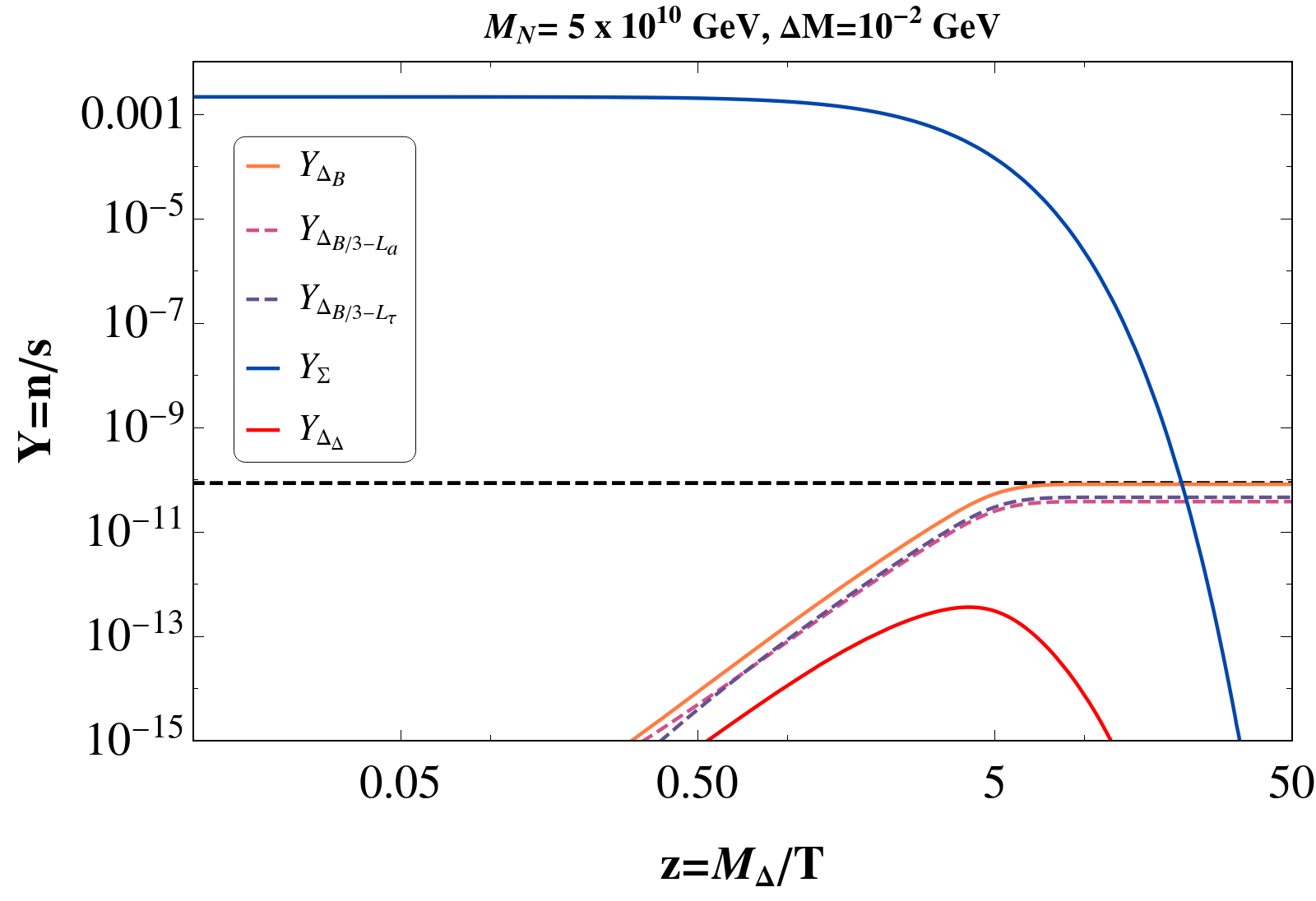}}~~
\subfigure[]{
	\includegraphics[scale=0.4]{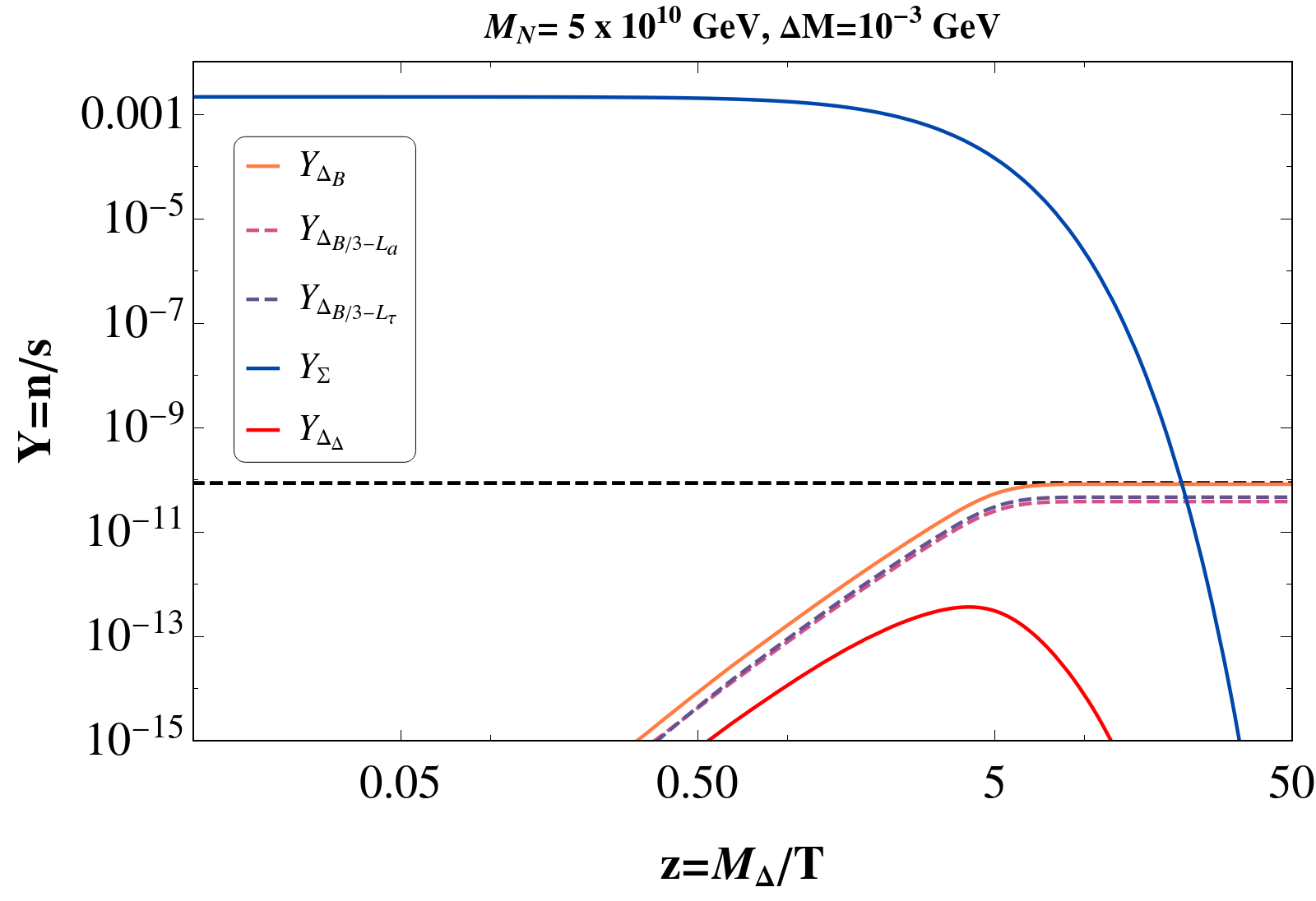}}~~

	\caption{Left panel depicts the evolution of the comoving number density of individual components of lepton asymmetry together with the overall baryon asymmetry for $M_{\Delta}=8.02 \times 10^9$ GeV and $\mu_2=1.14 \times 10^7 $ GeV. Right panel displays the same but for $M_{\Delta}=2.01 \times 10^9$ GeV and $\mu_2=1.48 \times 10^6$ GeV. Evolution of comoving number density of $\Sigma$, $\Delta_\Delta$ are also shown on both the plots. The values of $M_{\D}$ and $\mu_2$ are obatined from the lowest allowed points of Fig.~\ref{fig5}(a) (indicating by $\bigstar$).}
	\label{fig:ind-cont}
\end{figure}

Hereafter, to show explicitly the various contributions of flavor in the evolution of $Y_{\Delta_{B/3 - L_i}}$ ($i = a, \tau$), we pick up  the lowest possible values of $M_{\Delta}$ and corresponding $\mu_2$ (for a fixed value of $\Delta M = 10^{-2}~\text{GeV}$ and $M_N= 5\times10^{10}~\text{GeV}$) from Fig.~\ref{fig5}(a). Then in Fig.~\ref{fig:ind-cont}(a), we show the evolution of number density to entropy ratio ($Y=n/s$) for individual component of lepton asymmetries as well as the total baryon asymmetry with respect to $z=M_{\Delta}/T$.  While plotting, we have assumed that initially $\Delta(\bar{\Delta})$ were in equilibrium so $\Delta_{\Delta}=0$ and there were no lepton asymmetry. 
While the dark blue curve shows the evolution of $\Sigma$, abundances of $B/3 - L_i$ asymmetries along individual flavor directions are shown in purple ($a$ direction) and violet ($\tau$ direction) dashed lines. The orange line stands for the evolution of the baryon asymmetry which asymptotically merges with the black dashed horizontal line indicative of the correct baryon asymmetry of the Universe{\footnote{In a recent work \cite{Lavignac:2015gpa}, authors have shown the importance of incorporating density matrix formalism to evaluate the baryon asymmetry for triplet leptogenesis, even beyond $T \gtrsim 10^{12}$ GeV. In this formalism, diagonal entries of the density matrix indicate asymmetry along each lepton flavor direction while off-diagonal entries represent quantum correlations between different flavors. Though this is the most general approach, we have found that inclusion of the off-diagonal entries can only change the final result by $20\%$ or less (corresponding to the values of parameters involved in producing the plots of Fig.~\ref{fig:ind-cont}) and hence neglected here.}. Note that the dark blue curve starts to fall around $z=1$ due to the out of equilibrium 
decay of the triplet(anti-triplet) to different channels which in turn is reflected in the increase of lepton asymmetry (purple 
and violet-dashed lines). Around $z=5$, the number density of the lepton asymmetry (in all directions) starts to saturate. The red curve shows the evolution of asymmetry generated in $\Delta$ and $\bar{\Delta}$ particles. Next in Fig.~\ref{fig:ind-cont}(b), the similar evolution of flavors ($i=a,\tau$) becomes prominent once we choose the lowest possible value of $M_{\D}$ corresponding to $\D M = 10^{-3}~\text{GeV}$ from Fig.~\ref{fig5}(a).

Finally, we elaborate on how the mass-splitting $\Delta M$ intervenes in different parts of the present work. Obviously, $\Delta M$ has its most important role in DM phenomenology. As shown in Fig.~\ref{IHD_plot}, a value of $\Delta M \sim \mathcal{O}$(1) GeV or less is appropriate for having DM relic satisfaction having mass below TeV. Such a value therefore serves as the upper limit of $\Delta M$ while it is bounded from below by $\sim 10^{-4}$ GeV from the constraint on inelastic scattering amplitude of DM with detector nuclei. Turning into the neutrino part, we notice from Fig.~\ref{Y-type-I} that in order to maintain the type-II dominance toward the neutrino mass, the maximum allowed value of neutrino Yukawa coupling $Y$ has to be reduced with the increase of $\Delta M$ for a specific choice of RHN mass $M_N$. On the other hand, a larger $Y$ (and hence smaller $\Delta M$) is favored from the point of view of enhancing the CP asymmetry with a specific $M_N$. Therefore a judicious choice has to be made for choosing $\Delta M$ which not only be responsible for type-II dominance but also remains small enough so as to allow sufficient CP asymmetry. Such a choice has to be further guided by its upper ($\sim \mathcal{O}$(1) GeV) and lower ($10^{-4}$ GeV) limits. With this entire viable range of $\Delta M$, masses of the RHNs are found to be in the regime $\sim 10^{(9-12)}$ GeV.

\section{Conclusion} \label{sec6}

In this work, we present a simple extension of the basic type-II seesaw ($i.e.$ with one $SU(2)_L$ triplet in addition to SM) scenario including an additional RHN and one IHD, which can accommodate neutrino mass, dark matter as well as capable of explaining the baryon asymmetry of the Universe via leptogenesis mechanism. The interesting part of the study is the involvement of the DM multiplet, along with the RHN, in the vertex correction of the triplet's decay to two leptons which can successfully produce the required amount of CP asymmetry in order to address the baryon asymmetry of the Universe. Although the decay of RHN can also produce lepton asymmetry in the present setup, we assume a specific 
mass hierarchy $M_{\D}<M_N$ and hence the asymmetry generated by the decay of the triplet is the effective one. We incorporate the flavor effects in this triplet leptogenesis study as we aim to lower the triplet mass as much as possible in view of its accessibility at the collider. We find it is possible to generate sufficient lepton asymmetry with $M_{\Delta}$ as low as $\sim 10^{8}$ GeV.

Turning to the neutrino side, where the dominant contribution to the light neutrino mass follows from the tiny vev of the triplet, there exists a radiative contribution too which we restrict to be negligible by choosing the associated Yukawa to be small enough. This consideration is related to the mass splitting involved in the IHD which plays a twofold role here. Firstly, the IHD as a DM results with a specific range of this mass splitting ($10^{-4} - {\mathcal{O}}(1)$ GeV). Secondly, a smaller mass splitting (hence a larger neutrino Yukawa) 
turns out to be preferable for generating sufficient CP asymmetry in this flavored leptogenesis framework. Since the lower limit of $\Delta M$ is somewhat governed by the inelastic direct detection bound, in a way it restricts the mass of the triplet within a certain range. 

On the other hand, due to the involvement of particles like $\Delta^{\pm}$, $\Delta^{\pm\pm}$, $H^{\pm}$ and $N_R$, the present setup is also subjected to the constraints coming from the lepton flavor violating decays like $\mu \to e \gamma $. Keeping this in our mind,  we  calculate the $Br(\mu \to e \gamma)$ and found it to be many orders of magnitude smaller than that of the present upper bound on it ($<4.2\times10^{-13}$ at  $90\%$ C.L \cite{TheMEG:2016wtm}).

\bibliographystyle{utphys}
\bibliography{ref.bib}

\providecommand{\href}[2]{#2}\begingroup\raggedright\begin{thebibliography}{10}

\bibitem{Fukuda:1998mi}
{\bfseries Super-Kamiokande} Collaboration, Y.~Fukuda {\em et~al.}, ``{Evidence
  for oscillation of atmospheric neutrinos},''
  \href{http://dx.doi.org/10.1103/PhysRevLett.81.1562}{{\em Phys. Rev. Lett.}
  {\bfseries 81} (1998) 1562--1567},
  \href{http://arxiv.org/abs/hep-ex/9807003}{{\ttfamily arXiv:hep-ex/9807003}}.

\bibitem{Ahmad:2002jz}
{\bfseries SNO} Collaboration, Q.~Ahmad {\em et~al.}, ``{Direct evidence for
  neutrino flavor transformation from neutral current interactions in the
  Sudbury Neutrino Observatory},''
  \href{http://dx.doi.org/10.1103/PhysRevLett.89.011301}{{\em Phys. Rev. Lett.}
  {\bfseries 89} (2002) 011301},
  \href{http://arxiv.org/abs/nucl-ex/0204008}{{\ttfamily
  arXiv:nucl-ex/0204008}}.

\bibitem{Eguchi:2002dm}
{\bfseries KamLAND} Collaboration, K.~Eguchi {\em et~al.}, ``{First results
  from KamLAND: Evidence for reactor anti-neutrino disappearance},''
  \href{http://dx.doi.org/10.1103/PhysRevLett.90.021802}{{\em Phys. Rev. Lett.}
  {\bfseries 90} (2003) 021802},
  \href{http://arxiv.org/abs/hep-ex/0212021}{{\ttfamily arXiv:hep-ex/0212021}}.

\bibitem{Ahn:2002up}
{\bfseries K2K} Collaboration, M.~Ahn {\em et~al.}, ``{Indications of neutrino
  oscillation in a 250 km long baseline experiment},''
  \href{http://dx.doi.org/10.1103/PhysRevLett.90.041801}{{\em Phys. Rev. Lett.}
  {\bfseries 90} (2003) 041801},
  \href{http://arxiv.org/abs/hep-ex/0212007}{{\ttfamily arXiv:hep-ex/0212007}}.

\bibitem{Riotto:1999yt}
A.~Riotto and M.~Trodden, ``{Recent progress in baryogenesis},''
  \href{http://dx.doi.org/10.1146/annurev.nucl.49.1.35}{{\em Ann. Rev. Nucl.
  Part. Sci.} {\bfseries 49} (1999) 35--75},
  \href{http://arxiv.org/abs/hep-ph/9901362}{{\ttfamily arXiv:hep-ph/9901362}}.

\bibitem{Dine:2003ax}
M.~Dine and A.~Kusenko, ``{The Origin of the matter - antimatter asymmetry},''
  \href{http://dx.doi.org/10.1103/RevModPhys.76.1}{{\em Rev. Mod. Phys.}
  {\bfseries 76} (2003) 1},
  \href{http://arxiv.org/abs/hep-ph/0303065}{{\ttfamily arXiv:hep-ph/0303065}}.

\bibitem{Minkowski:1977sc}
P.~Minkowski, ``{$\mu \to e\gamma$ at a Rate of One Out of $10^{9}$ Muon
  Decays?},'' \href{http://dx.doi.org/10.1016/0370-2693(77)90435-X}{{\em Phys.
  Lett. B} {\bfseries 67} (1977) 421--428}.

\bibitem{GellMann:1980vs}
M.~Gell-Mann, P.~Ramond, and R.~Slansky, ``{Complex Spinors and Unified
  Theories},'' {\em Conf. Proc. C} {\bfseries 790927} (1979) 315--321,
  \href{http://arxiv.org/abs/1306.4669}{{\ttfamily arXiv:1306.4669 [hep-th]}}.

\bibitem{Mohapatra:1979ia}
R.~N. Mohapatra and G.~Senjanovic, ``{Neutrino Mass and Spontaneous Parity
  Nonconservation},'' \href{http://dx.doi.org/10.1103/PhysRevLett.44.912}{{\em
  Phys. Rev. Lett.} {\bfseries 44} (1980) 912}.

\bibitem{Schechter:1980gr}
J.~Schechter and J.~W.~F. Valle, ``{Neutrino Masses in SU(2) x U(1)
  Theories},'' \href{http://dx.doi.org/10.1103/PhysRevD.22.2227}{{\em Phys.
  Rev. D} {\bfseries 22} (1980) 2227}.

\bibitem{Fukugita:1986hr}
M.~Fukugita and T.~Yanagida, ``{Baryogenesis Without Grand Unification},''
  \href{http://dx.doi.org/10.1016/0370-2693(86)91126-3}{{\em Phys. Lett. B}
  {\bfseries 174} (1986) 45--47}.

\bibitem{Buchmuller:2004nz}
W.~Buchmuller, P.~Di~Bari, and M.~Plumacher, ``{Leptogenesis for
  pedestrians},'' \href{http://dx.doi.org/10.1016/j.aop.2004.02.003}{{\em
  Annals Phys.} {\bfseries 315} (2005) 305--351},
  \href{http://arxiv.org/abs/hep-ph/0401240}{{\ttfamily arXiv:hep-ph/0401240}}.

\bibitem{Anisimov:2007mw}
A.~Anisimov, S.~Blanchet, and P.~Di~Bari, ``{Viability of Dirac phase
  leptogenesis},'' \href{http://dx.doi.org/10.1088/1475-7516/2008/04/033}{{\em
  JCAP} {\bfseries 04} (2008) 033},
  \href{http://arxiv.org/abs/0707.3024}{{\ttfamily arXiv:0707.3024 [hep-ph]}}.

\bibitem{Davidson:2008bu}
S.~Davidson, E.~Nardi, and Y.~Nir, ``{Leptogenesis},''
  \href{http://dx.doi.org/10.1016/j.physrep.2008.06.002}{{\em Phys. Rept.}
  {\bfseries 466} (2008) 105--177},
  \href{http://arxiv.org/abs/0802.2962}{{\ttfamily arXiv:0802.2962 [hep-ph]}}.

\bibitem{Buchmuller:2005eh}
W.~Buchmuller, R.~Peccei, and T.~Yanagida, ``{Leptogenesis as the origin of
  matter},''
  \href{http://dx.doi.org/10.1146/annurev.nucl.55.090704.151558}{{\em Ann. Rev.
  Nucl. Part. Sci.} {\bfseries 55} (2005) 311--355},
  \href{http://arxiv.org/abs/hep-ph/0502169}{{\ttfamily arXiv:hep-ph/0502169}}.

\bibitem{Davoudiasl:2015jja}
H.~Davoudiasl and Y.~Zhang, ``{Baryon Number Violation via Majorana Neutrinos
  in the Early Universe, at the LHC, and Deep Underground},''
  \href{http://dx.doi.org/10.1103/PhysRevD.92.016005}{{\em Phys. Rev. D}
  {\bfseries 92} no.~1, (2015) 016005},
  \href{http://arxiv.org/abs/1504.07244}{{\ttfamily arXiv:1504.07244
  [hep-ph]}}.

\bibitem{Narendra:2017uxl}
N.~Narendra, N.~Sahoo, and N.~Sahu, ``{Dark matter assisted Dirac leptogenesis
  and neutrino mass},''
  \href{http://dx.doi.org/10.1016/j.nuclphysb.2018.09.007}{{\em Nucl. Phys. B}
  {\bfseries 936} (2018) 76--90},
  \href{http://arxiv.org/abs/1712.02960}{{\ttfamily arXiv:1712.02960
  [hep-ph]}}.

\bibitem{Dolan:2018qpy}
M.~J. Dolan, T.~P. Dutka, and R.~R. Volkas, ``{Dirac-Phase Thermal Leptogenesis
  in the extended Type-I Seesaw Model},''
  \href{http://dx.doi.org/10.1088/1475-7516/2018/06/012}{{\em JCAP} {\bfseries
  06} (2018) 012}, \href{http://arxiv.org/abs/1802.08373}{{\ttfamily
  arXiv:1802.08373 [hep-ph]}}.

\bibitem{Kashiwase:2012xd}
S.~Kashiwase and D.~Suematsu, ``{Baryon number asymmetry and dark matter in the
  neutrino mass model with an inert doublet},''
  \href{http://dx.doi.org/10.1103/PhysRevD.86.053001}{{\em Phys. Rev. D}
  {\bfseries 86} (2012) 053001},
  \href{http://arxiv.org/abs/1207.2594}{{\ttfamily arXiv:1207.2594 [hep-ph]}}.

\bibitem{Konar:2020vuu}
P.~Konar, A.~Mukherjee, A.~K. Saha, and S.~Show, ``{A dark clue to seesaw and
  leptogenesis in a pseudo-Dirac singlet doublet scenario with (non)standard
  cosmology},'' \href{http://dx.doi.org/10.1007/JHEP03(2021)044}{{\em JHEP}
  {\bfseries 03} (2021) 044}, \href{http://arxiv.org/abs/2007.15608}{{\ttfamily
  arXiv:2007.15608 [hep-ph]}}.

\bibitem{Barman:2021tgt}
B.~Barman, D.~Borah, and R.~Roshan, ``{Nonthermal leptogenesis and UV freeze-in
  of dark matter: Impact of inflationary reheating},''
  \href{http://dx.doi.org/10.1103/PhysRevD.104.035022}{{\em Phys. Rev. D}
  {\bfseries 104} no.~3, (2021) 035022},
  \href{http://arxiv.org/abs/2103.01675}{{\ttfamily arXiv:2103.01675
  [hep-ph]}}.

\bibitem{Bhattacharya:2021jli}
S.~Bhattacharya, R.~Roshan, A.~Sil, and D.~Vatsyayan, ``{Symmetry origin of
  Baryon Asymmetry, Dark Matter and Neutrino Mass},''
  \href{http://arxiv.org/abs/2105.06189}{{\ttfamily arXiv:2105.06189
  [hep-ph]}}.

\bibitem{Mohapatra:1980yp}
R.~N. Mohapatra and G.~Senjanovic, ``{Neutrino Masses and Mixings in Gauge
  Models with Spontaneous Parity Violation},''
  \href{http://dx.doi.org/10.1103/PhysRevD.23.165}{{\em Phys. Rev. D}
  {\bfseries 23} (1981) 165}.

\bibitem{Lazarides:1980nt}
G.~Lazarides, Q.~Shafi, and C.~Wetterich, ``{Proton Lifetime and Fermion Masses
  in an SO(10) Model},''
  \href{http://dx.doi.org/10.1016/0550-3213(81)90354-0}{{\em Nucl. Phys. B}
  {\bfseries 181} (1981) 287--300}.

\bibitem{Wetterich:1981bx}
C.~Wetterich, ``{Neutrino Masses and the Scale of B-L Violation},''
  \href{http://dx.doi.org/10.1016/0550-3213(81)90279-0}{{\em Nucl. Phys. B}
  {\bfseries 187} (1981) 343--375}.

\bibitem{Schechter:1981cv}
J.~Schechter and J.~W.~F. Valle, ``{Neutrino Decay and Spontaneous Violation of
  Lepton Number},'' \href{http://dx.doi.org/10.1103/PhysRevD.25.774}{{\em Phys.
  Rev. D} {\bfseries 25} (1982) 774}.

\bibitem{Brahmachari:1997cq}
B.~Brahmachari and R.~N. Mohapatra, ``{Unified explanation of the solar and
  atmospheric neutrino puzzles in a minimal supersymmetric SO(10) model},''
  \href{http://dx.doi.org/10.1103/PhysRevD.58.015001}{{\em Phys. Rev. D}
  {\bfseries 58} (1998) 015001},
  \href{http://arxiv.org/abs/hep-ph/9710371}{{\ttfamily arXiv:hep-ph/9710371}}.

\bibitem{Ma:1998dx}
E.~Ma and U.~Sarkar, ``{Neutrino masses and leptogenesis with heavy Higgs
  triplets},'' \href{http://dx.doi.org/10.1103/PhysRevLett.80.5716}{{\em Phys.
  Rev. Lett.} {\bfseries 80} (1998) 5716--5719},
  \href{http://arxiv.org/abs/hep-ph/9802445}{{\ttfamily arXiv:hep-ph/9802445}}.

\bibitem{Senami:2003jn}
M.~Senami and K.~Yamamoto, ``{Leptogenesis with supersymmetric Higgs triplets
  in TeV region},'' \href{http://dx.doi.org/10.1142/S0217751X06029478}{{\em
  Int. J. Mod. Phys. A} {\bfseries 21} (2006) 1291--1306},
  \href{http://arxiv.org/abs/hep-ph/0305202}{{\ttfamily arXiv:hep-ph/0305202}}.

\bibitem{GonzalezFelipe:2013jkc}
R.~Gonzalez~Felipe, F.~R. Joaquim, and H.~Serodio, ``{Flavoured CP asymmetries
  for type II seesaw leptogenesis},''
  \href{http://dx.doi.org/10.1142/S0217751X13501650}{{\em Int. J. Mod. Phys. A}
  {\bfseries 28} (2013) 1350165},
  \href{http://arxiv.org/abs/1301.0288}{{\ttfamily arXiv:1301.0288 [hep-ph]}}.

\bibitem{Lavignac:2015gpa}
S.~Lavignac and B.~Schmauch, ``{Flavour always matters in scalar triplet
  leptogenesis},'' \href{http://dx.doi.org/10.1007/JHEP05(2015)124}{{\em JHEP}
  {\bfseries 05} (2015) 124}, \href{http://arxiv.org/abs/1503.00629}{{\ttfamily
  arXiv:1503.00629 [hep-ph]}}.

\bibitem{Hambye:2003ka}
T.~Hambye and G.~Senjanovic, ``{Consequences of triplet seesaw for
  leptogenesis},'' \href{http://dx.doi.org/10.1016/j.physletb.2003.11.061}{{\em
  Phys. Lett. B} {\bfseries 582} (2004) 73--81},
  \href{http://arxiv.org/abs/hep-ph/0307237}{{\ttfamily arXiv:hep-ph/0307237}}.

\bibitem{Hambye:2005tk}
T.~Hambye, M.~Raidal, and A.~Strumia, ``{Efficiency and maximal CP-asymmetry of
  scalar triplet leptogenesis},''
  \href{http://dx.doi.org/10.1016/j.physletb.2005.11.007}{{\em Phys. Lett. B}
  {\bfseries 632} (2006) 667--674},
  \href{http://arxiv.org/abs/hep-ph/0510008}{{\ttfamily arXiv:hep-ph/0510008}}.

\bibitem{Hambye:2012fh}
T.~Hambye, ``{Leptogenesis: beyond the minimal type I seesaw scenario},''
  \href{http://dx.doi.org/10.1088/1367-2630/14/12/125014}{{\em New J. Phys.}
  {\bfseries 14} (2012) 125014},
  \href{http://arxiv.org/abs/1212.2888}{{\ttfamily arXiv:1212.2888 [hep-ph]}}.

\bibitem{AristizabalSierra:2014nzr}
D.~Aristizabal~Sierra, M.~Dhen, and T.~Hambye, ``{Scalar triplet flavored
  leptogenesis: a systematic approach},''
  \href{http://dx.doi.org/10.1088/1475-7516/2014/08/003}{{\em JCAP} {\bfseries
  08} (2014) 003}, \href{http://arxiv.org/abs/1401.4347}{{\ttfamily
  arXiv:1401.4347 [hep-ph]}}.

\bibitem{Mishra:2019sye}
S.~Mishra and A.~Giri, ``{Scalar triplet leptogenesis in the presence of
  right-handed neutrinos with $S_3$ symmetry},''
  \href{http://dx.doi.org/10.1088/1361-6471/ab7a86}{{\em J. Phys. G} {\bfseries
  47} no.~5, (2020) 055008}, \href{http://arxiv.org/abs/1909.12147}{{\ttfamily
  arXiv:1909.12147 [hep-ph]}}.

\bibitem{Rink:2020uvt}
T.~Rink, W.~Rodejohann, and K.~Schmitz, ``{Leptogenesis and low-energy CP
  violation in a type-II-dominated left-right seesaw model},''
  \href{http://dx.doi.org/10.1016/j.nuclphysb.2021.115552}{{\em Nucl. Phys. B}
  {\bfseries 972} (2021) 115552},
  \href{http://arxiv.org/abs/2006.03021}{{\ttfamily arXiv:2006.03021
  [hep-ph]}}.

\bibitem{Julian:1967zz}
W.~H. Julian, ``{On the Effect of Interstellar Material on Stellar Non-Circular
  Velocities in Disk Galaxies},'' \href{http://dx.doi.org/10.1086/149134}{{\em
  Astrophys. J.} {\bfseries 148} (1967) 175}.

\bibitem{Tegmark:2003ud}
{\bfseries SDSS} Collaboration, M.~Tegmark {\em et~al.}, ``{Cosmological
  parameters from SDSS and WMAP},''
  \href{http://dx.doi.org/10.1103/PhysRevD.69.103501}{{\em Phys. Rev. D}
  {\bfseries 69} (2004) 103501},
  \href{http://arxiv.org/abs/astro-ph/0310723}{{\ttfamily
  arXiv:astro-ph/0310723}}.

\bibitem{Bennett:2012zja}
{\bfseries WMAP} Collaboration, C.~L. Bennett {\em et~al.}, ``{Nine-Year
  Wilkinson Microwave Anisotropy Probe (WMAP) Observations: Final Maps and
  Results},'' \href{http://dx.doi.org/10.1088/0067-0049/208/2/20}{{\em
  Astrophys. J. Suppl.} {\bfseries 208} (2013) 20},
  \href{http://arxiv.org/abs/1212.5225}{{\ttfamily arXiv:1212.5225
  [astro-ph.CO]}}.

\bibitem{Clowe:2006eq}
D.~Clowe, M.~Bradac, A.~H. Gonzalez, M.~Markevitch, S.~W. Randall, C.~Jones,
  and D.~Zaritsky, ``{A direct empirical proof of the existence of dark
  matter},'' \href{http://dx.doi.org/10.1086/508162}{{\em Astrophys. J. Lett.}
  {\bfseries 648} (2006) L109--L113},
  \href{http://arxiv.org/abs/astro-ph/0608407}{{\ttfamily
  arXiv:astro-ph/0608407}}.

\bibitem{Datta:2021elq}
A.~Datta, R.~Roshan, and A.~Sil, ``{Imprint of the seesaw mechanism on feebly
  interacting dark matter and the baryon asymmetry},''
  \href{http://arxiv.org/abs/2104.02030}{{\ttfamily arXiv:2104.02030
  [hep-ph]}}.

\bibitem{LopezHonorez:2006gr}
L.~Lopez~Honorez, E.~Nezri, J.~F. Oliver, and M.~H.~G. Tytgat, ``{The Inert
  Doublet Model: An Archetype for Dark Matter},''
  \href{http://dx.doi.org/10.1088/1475-7516/2007/02/028}{{\em JCAP} {\bfseries
  0702} (2007) 028},
\href{http://arxiv.org/abs/hep-ph/0612275}{{\ttfamily arXiv:hep-ph/0612275
  [hep-ph]}}.
%%CITATION = HEP-PH/0612275;%%.

\bibitem{Honorez:2010re}
L.~Lopez~Honorez and C.~E. Yaguna, ``{The inert doublet model of dark matter
  revisited},'' \href{http://dx.doi.org/10.1007/JHEP09(2010)046}{{\em JHEP}
  {\bfseries 09} (2010) 046},
\href{http://arxiv.org/abs/1003.3125}{{\ttfamily arXiv:1003.3125 [hep-ph]}}.
%%CITATION = ARXIV:1003.3125;%%.

\bibitem{Belyaev:2016lok}
A.~Belyaev, G.~Cacciapaglia, I.~P. Ivanov, F.~Rojas-Abatte, and M.~Thomas,
  ``{Anatomy of the Inert Two Higgs Doublet Model in the light of the LHC and
  non-LHC Dark Matter Searches},''
  \href{http://dx.doi.org/10.1103/PhysRevD.97.035011}{{\em Phys. Rev.}
  {\bfseries D97} no.~3, (2018) 035011},
\href{http://arxiv.org/abs/1612.00511}{{\ttfamily arXiv:1612.00511 [hep-ph]}}.
%%CITATION = ARXIV:1612.00511;%%.

\bibitem{Choubey:2017hsq}
S.~Choubey and A.~Kumar, ``{Inflation and Dark Matter in the Inert Doublet
  Model},'' \href{http://dx.doi.org/10.1007/JHEP11(2017)080}{{\em JHEP}
  {\bfseries 11} (2017) 080},
\href{http://arxiv.org/abs/1707.06587}{{\ttfamily arXiv:1707.06587 [hep-ph]}}.
%%CITATION = ARXIV:1707.06587;%%.

\bibitem{LopezHonorez:2010tb}
L.~Lopez~Honorez and C.~E. Yaguna, ``{A new viable region of the inert doublet
  model},'' \href{http://dx.doi.org/10.1088/1475-7516/2011/01/002}{{\em JCAP}
  {\bfseries 1101} (2011) 002},
\href{http://arxiv.org/abs/1011.1411}{{\ttfamily arXiv:1011.1411 [hep-ph]}}.
%%CITATION = ARXIV:1011.1411;%%.

\bibitem{Ilnicka:2015jba}
A.~Ilnicka, M.~Krawczyk, and T.~Robens, ``{Inert Doublet Model in light of LHC
  Run I and astrophysical data},''
  \href{http://dx.doi.org/10.1103/PhysRevD.93.055026}{{\em Phys. Rev.}
  {\bfseries D93} no.~5, (2016) 055026},
\href{http://arxiv.org/abs/1508.01671}{{\ttfamily arXiv:1508.01671 [hep-ph]}}.
%%CITATION = ARXIV:1508.01671;%%.

\bibitem{Arhrib:2013ela}
A.~Arhrib, Y.-L.~S. Tsai, Q.~Yuan, and T.-C. Yuan, ``{An Updated Analysis of
  Inert Higgs Doublet Model in light of the Recent Results from LUX, PLANCK,
  AMS-02 and LHC},''
  \href{http://dx.doi.org/10.1088/1475-7516/2014/06/030}{{\em JCAP} {\bfseries
  1406} (2014) 030},
\href{http://arxiv.org/abs/1310.0358}{{\ttfamily arXiv:1310.0358 [hep-ph]}}.
%%CITATION = ARXIV:1310.0358;%%.

\bibitem{Cao:2007rm}
Q.-H. Cao, E.~Ma, and G.~Rajasekaran, ``{Observing the Dark Scalar Doublet and
  its Impact on the Standard-Model Higgs Boson at Colliders},''
  \href{http://dx.doi.org/10.1103/PhysRevD.76.095011}{{\em Phys. Rev.}
  {\bfseries D76} (2007) 095011},
\href{http://arxiv.org/abs/0708.2939}{{\ttfamily arXiv:0708.2939 [hep-ph]}}.
%%CITATION = ARXIV:0708.2939;%%.

\bibitem{Lundstrom:2008ai}
E.~Lundstrom, M.~Gustafsson, and J.~Edsjo, ``{The Inert Doublet Model and LEP
  II Limits},'' \href{http://dx.doi.org/10.1103/PhysRevD.79.035013}{{\em Phys.
  Rev.} {\bfseries D79} (2009) 035013},
\href{http://arxiv.org/abs/0810.3924}{{\ttfamily arXiv:0810.3924 [hep-ph]}}.
%%CITATION = ARXIV:0810.3924;%%.

\bibitem{Gustafsson:2012aj}
M.~Gustafsson, S.~Rydbeck, L.~Lopez-Honorez, and E.~Lundstrom, ``{Status of the
  Inert Doublet Model and the Role of multileptons at the LHC},''
  \href{http://dx.doi.org/10.1103/PhysRevD.86.075019}{{\em Phys. Rev.}
  {\bfseries D86} (2012) 075019},
\href{http://arxiv.org/abs/1206.6316}{{\ttfamily arXiv:1206.6316 [hep-ph]}}.
%%CITATION = ARXIV:1206.6316;%%.

\bibitem{Borah:2017dfn}
D.~Borah and A.~Gupta, ``{New viable region of an inert Higgs doublet dark
  matter model with scotogenic extension},''
  \href{http://dx.doi.org/10.1103/PhysRevD.96.115012}{{\em Phys. Rev. D}
  {\bfseries 96} no.~11, (2017) 115012},
  \href{http://arxiv.org/abs/1706.05034}{{\ttfamily arXiv:1706.05034
  [hep-ph]}}.

\bibitem{Kalinowski:2018ylg}
J.~Kalinowski, W.~Kotlarski, T.~Robens, D.~Sokolowska, and A.~F. Zarnecki,
  ``{Benchmarking the Inert Doublet Model for $e^+ e^-$ colliders},''
  \href{http://dx.doi.org/10.1007/JHEP12(2018)081}{{\em JHEP} {\bfseries 12}
  (2018) 081},
\href{http://arxiv.org/abs/1809.07712}{{\ttfamily arXiv:1809.07712 [hep-ph]}}.
%%CITATION = ARXIV:1809.07712;%%.

\bibitem{Bhardwaj:2019mts}
A.~Bhardwaj, P.~Konar, T.~Mandal, and S.~Sadhukhan, ``{Probing Inert Doublet
  Model using jet substructure with multivariate analysis},''
\href{http://arxiv.org/abs/1905.04195}{{\ttfamily arXiv:1905.04195 [hep-ph]}}.
%%CITATION = ARXIV:1905.04195;%%.

\bibitem{Borah:2019aeq}
D.~Borah, R.~Roshan, and A.~Sil, ``{Minimal two-component scalar doublet dark
  matter with radiative neutrino mass},''
  \href{http://dx.doi.org/10.1103/PhysRevD.100.055027}{{\em Phys. Rev. D}
  {\bfseries 100} no.~5, (2019) 055027},
  \href{http://arxiv.org/abs/1904.04837}{{\ttfamily arXiv:1904.04837
  [hep-ph]}}.

\bibitem{Bhattacharya:2019fgs}
S.~Bhattacharya, P.~Ghosh, A.~K. Saha, and A.~Sil, ``{Two component dark matter
  with inert Higgs doublet: neutrino mass, high scale validity and collider
  searches},'' \href{http://dx.doi.org/10.1007/JHEP03(2020)090}{{\em JHEP}
  {\bfseries 03} (2020) 090}, \href{http://arxiv.org/abs/1905.12583}{{\ttfamily
  arXiv:1905.12583 [hep-ph]}}.

\bibitem{Bhattacharya:2019tqq}
S.~Bhattacharya, N.~Chakrabarty, R.~Roshan, and A.~Sil, ``{Multicomponent dark
  matter in extended $U(1)_{B-L}$: neutrino mass and high scale validity},''
  \href{http://dx.doi.org/10.1088/1475-7516/2020/04/013}{{\em JCAP} {\bfseries
  04} (2020) 013}, \href{http://arxiv.org/abs/1910.00612}{{\ttfamily
  arXiv:1910.00612 [hep-ph]}}.

\bibitem{Chakrabarty:2021kmr}
N.~Chakrabarty, R.~Roshan, and A.~Sil, ``{Two Component Doublet-Triplet Scalar
  Dark Matter stabilising the Electroweak vacuum},''
  \href{http://arxiv.org/abs/2102.06032}{{\ttfamily arXiv:2102.06032
  [hep-ph]}}.

\bibitem{Abada:2006fw}
A.~Abada, S.~Davidson, F.-X. Josse-Michaux, M.~Losada, and A.~Riotto, ``{Flavor
  issues in leptogenesis},''
  \href{http://dx.doi.org/10.1088/1475-7516/2006/04/004}{{\em JCAP} {\bfseries
  04} (2006) 004}, \href{http://arxiv.org/abs/hep-ph/0601083}{{\ttfamily
  arXiv:hep-ph/0601083}}.

\bibitem{Nardi:2006fx}
E.~Nardi, Y.~Nir, E.~Roulet, and J.~Racker, ``{The Importance of flavor in
  leptogenesis},'' \href{http://dx.doi.org/10.1088/1126-6708/2006/01/164}{{\em
  JHEP} {\bfseries 01} (2006) 164},
  \href{http://arxiv.org/abs/hep-ph/0601084}{{\ttfamily arXiv:hep-ph/0601084}}.

\bibitem{Blanchet:2006be}
S.~Blanchet and P.~Di~Bari, ``{Flavor effects on leptogenesis predictions},''
  \href{http://dx.doi.org/10.1088/1475-7516/2007/03/018}{{\em JCAP} {\bfseries
  03} (2007) 018}, \href{http://arxiv.org/abs/hep-ph/0607330}{{\ttfamily
  arXiv:hep-ph/0607330}}.

\bibitem{Dev:2017trv}
P.~S.~B. Dev, P.~Di~Bari, B.~Garbrecht, S.~Lavignac, P.~Millington, and
  D.~Teresi, ``{Flavor effects in leptogenesis},''
  \href{http://dx.doi.org/10.1142/S0217751X18420010}{{\em Int. J. Mod. Phys. A}
  {\bfseries 33} (2018) 1842001},
  \href{http://arxiv.org/abs/1711.02861}{{\ttfamily arXiv:1711.02861
  [hep-ph]}}.

\bibitem{Datta:2021zzf}
A.~Datta, B.~Karmakar, and A.~Sil, ``{Flavored Leptogenesis and Neutrino Mass
  with $A_4$ Symmetry},'' \href{http://arxiv.org/abs/2106.06773}{{\ttfamily
  arXiv:2106.06773 [hep-ph]}}.

\bibitem{deFlorian:2016spz}
{\bfseries LHC Higgs Cross Section Working Group} Collaboration, D.~de~Florian
  {\em et~al.}, ``{Handbook of LHC Higgs Cross Sections: 4. Deciphering the
  Nature of the Higgs Sector},''
\href{http://arxiv.org/abs/1610.07922}{{\ttfamily arXiv:1610.07922 [hep-ph]}}.
%%CITATION = ARXIV:1610.07922;%%.

\bibitem{ParticleDataGroup:2020ssz}
{\bfseries Particle Data Group} Collaboration, P.~A. Zyla {\em et~al.},
  ``{Review of Particle Physics},''
  \href{http://dx.doi.org/10.1093/ptep/ptaa104}{{\em PTEP} {\bfseries 2020}
  no.~8, (2020) 083C01}.

\bibitem{Sirunyan:2018koj}
{\bfseries CMS} Collaboration, A.~M. Sirunyan {\em et~al.}, ``{Combined
  measurements of Higgs boson couplings in proton–proton collisions at
  $\sqrt{s}=13\,\text {Te}\text {V} $},''
  \href{http://dx.doi.org/10.1140/epjc/s10052-019-6909-y}{{\em Eur. Phys. J.}
  {\bfseries C79} no.~5, (2019) 421},
\href{http://arxiv.org/abs/1809.10733}{{\ttfamily arXiv:1809.10733 [hep-ex]}}.
%%CITATION = ARXIV:1809.10733;%%.

\bibitem{Aaboud:2018xdt}
{\bfseries ATLAS} Collaboration, M.~Aaboud {\em et~al.}, ``{Measurements of
  Higgs boson properties in the diphoton decay channel with 36 fb$^{-1}$ of
  $pp$ collision data at $\sqrt{s} = 13$ TeV with the ATLAS detector},''
  \href{http://dx.doi.org/10.1103/PhysRevD.98.052005}{{\em Phys. Rev.}
  {\bfseries D98} (2018) 052005},
\href{http://arxiv.org/abs/1802.04146}{{\ttfamily arXiv:1802.04146 [hep-ex]}}.
%%CITATION = ARXIV:1802.04146;%%.

\bibitem{deSalas:2020pgw}
P.~F. de~Salas, D.~V. Forero, S.~Gariazzo, P.~Mart\'\i{}nez-Mirav\'e, O.~Mena,
  C.~A. Ternes, M.~T\'ortola, and J.~W.~F. Valle, ``{2020 global reassessment
  of the neutrino oscillation picture},''
  \href{http://dx.doi.org/10.1007/JHEP02(2021)071}{{\em JHEP} {\bfseries 02}
  (2021) 071}, \href{http://arxiv.org/abs/2006.11237}{{\ttfamily
  arXiv:2006.11237 [hep-ph]}}.

\bibitem{Ahriche:2017iar}
A.~Ahriche, A.~Jueid, and S.~Nasri, ``{Radiative neutrino mass and Majorana
  dark matter within an inert Higgs doublet model},''
  \href{http://dx.doi.org/10.1103/PhysRevD.97.095012}{{\em Phys. Rev. D}
  {\bfseries 97} no.~9, (2018) 095012},
  \href{http://arxiv.org/abs/1710.03824}{{\ttfamily arXiv:1710.03824
  [hep-ph]}}.

\bibitem{Gondolo:1990dk}
P.~Gondolo and G.~Gelmini, ``{Cosmic abundances of stable particles: Improved
  analysis},'' \href{http://dx.doi.org/10.1016/0550-3213(91)90438-4}{{\em Nucl.
  Phys. B} {\bfseries 360} (1991) 145--179}.

\bibitem{Barducci:2016pcb}
D.~Barducci, G.~Belanger, J.~Bernon, F.~Boudjema, J.~Da~Silva, S.~Kraml,
  U.~Laa, and A.~Pukhov, ``{Collider limits on new physics within
  micrOMEGAs$\_$4.3}'' \href{http://dx.doi.org/10.1016/j.cpc.2017.08.028}{{\em
  Comput. Phys. Commun.} {\bfseries 222} (2018) 327--338},
  \href{http://arxiv.org/abs/1606.03834}{{\ttfamily arXiv:1606.03834
  [hep-ph]}}.

\bibitem{DuttaBanik:2020vfr}
A.~Dutta~Banik, R.~Roshan, and A.~Sil, ``{Neutrino mass and asymmetric dark
  matter: study with inert Higgs doublet and high scale validity},''
  \href{http://dx.doi.org/10.1088/1475-7516/2021/03/037}{{\em JCAP} {\bfseries
  03} (2021) 037}, \href{http://arxiv.org/abs/2011.04371}{{\ttfamily
  arXiv:2011.04371 [hep-ph]}}.

\bibitem{Akerib:2016vxi}
{\bfseries LUX} Collaboration, D.~S. Akerib {\em et~al.}, ``{Results from a
  search for dark matter in the complete LUX exposure},''
  \href{http://dx.doi.org/10.1103/PhysRevLett.118.021303}{{\em Phys. Rev.
  Lett.} {\bfseries 118} no.~2, (2017) 021303},
\href{http://arxiv.org/abs/1608.07648}{{\ttfamily arXiv:1608.07648
  [astro-ph.CO]}}.
%%CITATION = ARXIV:1608.07648;%%.

\bibitem{Tan:2016zwf}
{\bfseries PandaX-II} Collaboration, A.~Tan {\em et~al.}, ``{Dark Matter
  Results from First 98.7 Days of Data from the PandaX-II Experiment},''
  \href{http://dx.doi.org/10.1103/PhysRevLett.117.121303}{{\em Phys. Rev.
  Lett.} {\bfseries 117} no.~12, (2016) 121303},
\href{http://arxiv.org/abs/1607.07400}{{\ttfamily arXiv:1607.07400 [hep-ex]}}.
%%CITATION = ARXIV:1607.07400;%%.

\bibitem{Cui:2017nnn}
{\bfseries PandaX-II} Collaboration, X.~Cui {\em et~al.}, ``{Dark Matter
  Results From 54-Ton-Day Exposure of PandaX-II Experiment},''
  \href{http://dx.doi.org/10.1103/PhysRevLett.119.181302}{{\em Phys. Rev.
  Lett.} {\bfseries 119} no.~18, (2017) 181302},
\href{http://arxiv.org/abs/1708.06917}{{\ttfamily arXiv:1708.06917
  [astro-ph.CO]}}.
%%CITATION = ARXIV:1708.06917;%%.

\bibitem{Aprile:2017iyp}
{\bfseries XENON} Collaboration, E.~Aprile {\em et~al.}, ``{First Dark Matter
  Search Results from the XENON1T Experiment},''
  \href{http://dx.doi.org/10.1103/PhysRevLett.119.181301}{{\em Phys. Rev.
  Lett.} {\bfseries 119} no.~18, (2017) 181301},
\href{http://arxiv.org/abs/1705.06655}{{\ttfamily arXiv:1705.06655
  [astro-ph.CO]}}.
%%CITATION = ARXIV:1705.06655;%%.

\bibitem{Aprile:2018dbl}
{\bfseries XENON} Collaboration, E.~Aprile {\em et~al.}, ``{Dark Matter Search
  Results from a One Ton-Year Exposure of XENON1T},''
  \href{http://dx.doi.org/10.1103/PhysRevLett.121.111302}{{\em Phys. Rev.
  Lett.} {\bfseries 121} no.~11, (2018) 111302},
\href{http://arxiv.org/abs/1805.12562}{{\ttfamily arXiv:1805.12562
  [astro-ph.CO]}}.
%%CITATION = ARXIV:1805.12562;%%.

\bibitem{Barbieri:2006dq}
R.~Barbieri, L.~J. Hall, and V.~S. Rychkov, ``{Improved naturalness with a
  heavy Higgs: An Alternative road to LHC physics},''
  \href{http://dx.doi.org/10.1103/PhysRevD.74.015007}{{\em Phys. Rev. D}
  {\bfseries 74} (2006) 015007},
  \href{http://arxiv.org/abs/hep-ph/0603188}{{\ttfamily arXiv:hep-ph/0603188}}.

\bibitem{Giedt:2009mr}
J.~Giedt, A.~W. Thomas, and R.~D. Young, ``{Dark matter, the CMSSM and lattice
  QCD},'' \href{http://dx.doi.org/10.1103/PhysRevLett.103.201802}{{\em Phys.
  Rev. Lett.} {\bfseries 103} (2009) 201802},
  \href{http://arxiv.org/abs/0907.4177}{{\ttfamily arXiv:0907.4177 [hep-ph]}}.

\bibitem{Cirelli:2009uv}
M.~Cirelli and A.~Strumia, ``{Minimal Dark Matter: Model and results},''
  \href{http://dx.doi.org/10.1088/1367-2630/11/10/105005}{{\em New J. Phys.}
  {\bfseries 11} (2009) 105005},
  \href{http://arxiv.org/abs/0903.3381}{{\ttfamily arXiv:0903.3381 [hep-ph]}}.

\bibitem{Arina:2009um}
C.~Arina, F.-S. Ling, and M.~H.~G. Tytgat, ``{IDM and iDM or The Inert Doublet
  Model and Inelastic Dark Matter},''
  \href{http://dx.doi.org/10.1088/1475-7516/2009/10/018}{{\em JCAP} {\bfseries
  10} (2009) 018}, \href{http://arxiv.org/abs/0907.0430}{{\ttfamily
  arXiv:0907.0430 [hep-ph]}}.

\bibitem{Eiteneuer:2017hoh}
B.~Eiteneuer, A.~Goudelis, and J.~Heisig, ``{The inert doublet model in the
  light of Fermi-LAT gamma-ray data: a global fit analysis},''
  \href{http://dx.doi.org/10.1140/epjc/s10052-017-5166-1}{{\em Eur. Phys. J. C}
  {\bfseries 77} no.~9, (2017) 624},
  \href{http://arxiv.org/abs/1705.01458}{{\ttfamily arXiv:1705.01458
  [hep-ph]}}.

\bibitem{MAGIC:2016xys}
{\bfseries MAGIC, Fermi-LAT} Collaboration, M.~L. Ahnen {\em et~al.}, ``{Limits
  to Dark Matter Annihilation Cross-Section from a Combined Analysis of MAGIC
  and Fermi-LAT Observations of Dwarf Satellite Galaxies},''
  \href{http://dx.doi.org/10.1088/1475-7516/2016/02/039}{{\em JCAP} {\bfseries
  02} (2016) 039}, \href{http://arxiv.org/abs/1601.06590}{{\ttfamily
  arXiv:1601.06590 [astro-ph.HE]}}.

\bibitem{Cyburt:2015mya}
R.~H. Cyburt, B.~D. Fields, K.~A. Olive, and T.-H. Yeh, ``{Big Bang
  Nucleosynthesis: 2015},''
  \href{http://dx.doi.org/10.1103/RevModPhys.88.015004}{{\em Rev. Mod. Phys.}
  {\bfseries 88} (2016) 015004},
  \href{http://arxiv.org/abs/1505.01076}{{\ttfamily arXiv:1505.01076
  [astro-ph.CO]}}.

\bibitem{Planck:2018vyg}
{\bfseries Planck} Collaboration, N.~Aghanim {\em et~al.}, ``{Planck 2018
  results. VI. Cosmological parameters},''
  \href{http://dx.doi.org/10.1051/0004-6361/201833910}{{\em Astron. Astrophys.}
  {\bfseries 641} (2020) A6}, \href{http://arxiv.org/abs/1807.06209}{{\ttfamily
  arXiv:1807.06209 [astro-ph.CO]}}. [Erratum: Astron.Astrophys. 652, C4
  (2021)].

\bibitem{TheMEG:2016wtm}
{\bfseries MEG} Collaboration, A.~M. Baldini {\em et~al.}, ``{Search for the
  lepton flavour violating decay $\mu ^+ \rightarrow \mathrm {e}^+ \gamma $
  with the full dataset of the MEG experiment},''
  \href{http://dx.doi.org/10.1140/epjc/s10052-016-4271-x}{{\em Eur. Phys. J. C}
  {\bfseries 76} no.~8, (2016) 434},
  \href{http://arxiv.org/abs/1605.05081}{{\ttfamily arXiv:1605.05081
  [hep-ex]}}.

\end{thebibliography}\endgroup
\end{document}